\documentclass[aps,prl,twocolumn,10pt,a4paper,superscriptaddress]{revtex4}
\pdfoutput=1

\usepackage{color}
\usepackage[latin1]{inputenc}
\usepackage{graphicx}
\usepackage{amsmath,amssymb}
\usepackage{dsfont}
\usepackage{hyperref}
\hypersetup{colorlinks=true,
						linkcolor=blue,
						anchorcolor=blue,
						citecolor=blue,
						filecolor=blue,
						menucolor=blue,
						urlcolor=blue
}

\renewcommand{\figurename}[1]{Fig.\;}
\makeatletter
\renewcommand\frontmatter@abstractwidth{\textwidth}
\makeatother

\begin{document}

\title{Realization of density-dependent Peierls phases to engineer quantized gauge fields coupled to ultracold matter}

\author{Frederik G\"org}
\affiliation{Institute for Quantum Electronics, ETH Zurich, 8093 Zurich, Switzerland}

\author{Kilian Sandholzer}
\affiliation{Institute for Quantum Electronics, ETH Zurich, 8093 Zurich, Switzerland}

\author{Joaqu\'in Minguzzi}
\affiliation{Institute for Quantum Electronics, ETH Zurich, 8093 Zurich, Switzerland}

\author{R\'emi Desbuquois}
\affiliation{Institute for Quantum Electronics, ETH Zurich, 8093 Zurich, Switzerland}

\author{Michael Messer}
\affiliation{Institute for Quantum Electronics, ETH Zurich, 8093 Zurich, Switzerland}

\author{Tilman Esslinger}
\affiliation{Institute for Quantum Electronics, ETH Zurich, 8093 Zurich, Switzerland}

\begin{abstract}
\textbf{
Gauge fields that appear in models of high-energy and condensed matter physics are dynamical quantum degrees of freedom due to their coupling to matter fields. Since the dynamics of these strongly correlated systems is hard to compute, it was proposed to implement this basic coupling mechanism in quantum simulation platforms with the ultimate goal to emulate lattice gauge theories. Here, we realize the fundamental ingredient for a density-dependent gauge field acting on ultracold fermions in an optical lattice by engineering non-trivial Peierls phases that depend on the site occupations. We propose and implement a Floquet scheme that relies on breaking time-reversal symmetry (TRS) by driving the lattice simultaneously at two frequencies which are resonant with the onsite interactions. This induces density-assisted tunnelling processes that are controllable in amplitude and phase. We demonstrate techniques in a Hubbard dimer to quantify the amplitude and to directly measure the Peierls phase with respect to the single-particle hopping. The tunnel coupling features two distinct regimes as a function of the modulation amplitudes, which can be characterised by a $\mathbb{Z}_2$-invariant. Moreover, we provide a full tomography of the winding structure of the Peierls phase around a Dirac point that appears in the driving parameter space.
}
\end{abstract}

\maketitle

The fundamental manifestation of a gauge field in electromagnetism is the Lorentz force acting on charged particles. 
In ultracold Bose and Fermi gases, the charge neutrality of the atoms requires to engineer synthetic magnetic fields \cite{Goldman2014a,Cooper2019}. 
This has been achieved for bulk systems by a rotation of the gas or a suitable coupling of momentum states via Raman lasers \cite{Cooper2008,Lin2009}. 
For a tight-binding model on a lattice, the equivalent of an Aharonov-Bohm phase can be synthesized with Peierls phases resulting from a complex-valued tunnelling matrix element. 
Such phases can be engineered in a Floquet approach by a suitable driving scheme \cite{Bukov2015,Eckardt2017}, which has been used in cold atom experiments to generate static gauge fields \cite{Aidelsburger2011,Miyake2013,Struck2013,Jotzu2014}. 
So far, these synthetic fields for atoms in optical lattices were intrinsically classical, as they did not feature a back-action from the atoms. 
In contrast, gauge fields appearing in nature are dynamical in the sense that they are influenced by the spatial configuration and motion of the matter field \cite{Cheng1991,Levin2005,Savary2017}. 
Therefore, as a first step towards the simulation of lattice gauge theories \cite{Wiese2013,Zohar2015,Dalmonte2016,Martinez2016}, it is necessary to implement a coupling mechanism between the gauge and matter fields. 
One possibility is to engineer density-dependent gauge fields by making use of interactions \cite{Edmonds2013a}. 
Such a scheme has recently been implemented experimentally by adding a directional mean-field shift in momentum space to a Bose-Einstein condensate \cite{Clark2018}. 
For tight-binding models a back-action mechanism encoded in Peierls phases that depend on the occupation of the lattice sites has been suggested theoretically \cite{Keilmann2011,Greschner2014,Greschner2015,Bermudez2015,Cardarelli2016,Strater2016,Barbiero2018}. 

We propose and experimentally realize a scheme based on a Floquet-engineering approach, which allows us to control both the amplitude and Peierls phase of density-dependent tunnelling matrix elements. In order to obtain non-trivial hopping phases, we explicitly break TRS by modulating the position of an optical lattice simultaneously at two frequencies $\omega/(2\pi)$ and $2\omega/(2\pi)$ \cite{Struck2012}. By tuning the Hubbard on-site interactions close to a resonance $U=l\hbar\omega$ ($l \in\mathbb{Z}$), we induce complex density-assisted tunnelling processes $t^{(l)}_{\text{eff}}=|t^{(l)}_{\text{eff}}|\exp (i\psi^{(l)})$ by exchanging photons with the drive (for realizations of real-valued density-assisted hoppings induced by a periodic modulation see Refs.\;\cite{Stoferle2004,Jordens2008,Greif2011,Chen2011,Ma2011,Meinert2016,Desbuquois2017,Gorg2018,Messer2018,Xu2018,Sandholzer2018,Schweizer2019}). Using Floquet theory, we derive an effective static Hamiltonian describing the long-term dynamics of the system \cite{Bukov2015} (see Methods and supplementary information (SI)), which is given by
\begin{equation}
\hat{H}^{(l)}_{\text{eff}}=-\sum_{\langle \textbf{i}, \textbf{j} \rangle,\sigma} \left(\hat{t}_{\langle \textbf{i}, \textbf{j} \rangle,\bar{\sigma}}^{(l)} e^{i\hat{\mathcal{A}}_{\langle \textbf{i}, \textbf{j} \rangle,\bar{\sigma}}^{(l)}} \hat{c}^{\dagger}_{\textbf{j}\sigma}\hat{c}_{\textbf{i}\sigma}+\text{h.c.}\right). 
\label{HeffDynGaugeFields}
\end{equation}
Here, the operators $\hat{c}^{\dagger}_{\textbf{j}\sigma}$ and $\hat{c}_{\textbf{j}\sigma}$ create and annihilate a fermion on site $\textbf{j}$ in spin state $\sigma\in\{\uparrow,\downarrow\}$, respectively. In general, both the tunneling amplitude $\hat{t}_{\langle \textbf{i}, \textbf{j} \rangle,\bar{\sigma}}^{(l)}$ and phase $\hat{\mathcal{A}}_{\langle \textbf{i}, \textbf{j} \rangle,\bar{\sigma}}^{(l)}$ for atoms in state $\sigma$ are operators, since they depend on the configuration of particles in the opposite state $\bar{\sigma}$, which therefore act as a link variable on the nearest-neighbor bond $\langle \textbf{i}, \textbf{j} \rangle$. Using a spin-1 representation in the eigenbasis of $\hat{\tau}_{\langle \textbf{i},\textbf{j} \rangle,\sigma}^z = \hat{n}_{\textbf{i}\sigma}-\hat{n}_{\textbf{j}\sigma}$ where $\hat{n}_{\textbf{i}\sigma}=\hat{c}^{\dagger}_{\textbf{i}\sigma}\hat{c}_{\textbf{i}\sigma}$ is the number operator, the tunneling operators can be expressed as $\hat{t}_{\langle \textbf{i}, \textbf{j} \rangle,\bar{\sigma}}^{(l)}=\text{diag}(|t_{\text{eff},\text{L}}^{(l)}|,|t_{\text{eff}}^{(0)}|,|t_{\text{eff},\text{R}}^{(l)}|)$ and $\hat{\mathcal{A}}_{\langle \textbf{i}, \textbf{j} \rangle,\bar{\sigma}}^{(l)}=\text{diag}(-\psi_{\text{L}}^{(l)},\psi^{(0)},\psi_{\text{R}}^{(l)})$, respectively (see SI). Here, $l=0$ corresponds to the single-particle hopping and R and L denote density-assisted tunnelings ($|l|>0$) that involve an atom in the opposite spin state on the right and left site, respectively. Importantly, all tunneling amplitudes $|t^{(l)}_{\text{eff}}|$ and Peierls phases $\psi^{(l)}$ can be tuned independently in our scheme via the driving strengths and phases. Since $\hat{\mathcal{A}}_{\langle \textbf{i}, \textbf{j} \rangle,\bar{\sigma}}^{(l)}$ is an operator, it acts as a dynamical gauge field for atoms in spin state $\sigma$. The Hamiltonian (\ref{HeffDynGaugeFields}) is reminiscent of a lattice gauge theory and we show in the SI how to engineer dynamical $\mathbb{Z}_2$ and $\mathbb{Z}_3$ gauge fields with a two-frequency driving scheme. 

\begin{figure}[htb]
	\includegraphics[width=89mm]{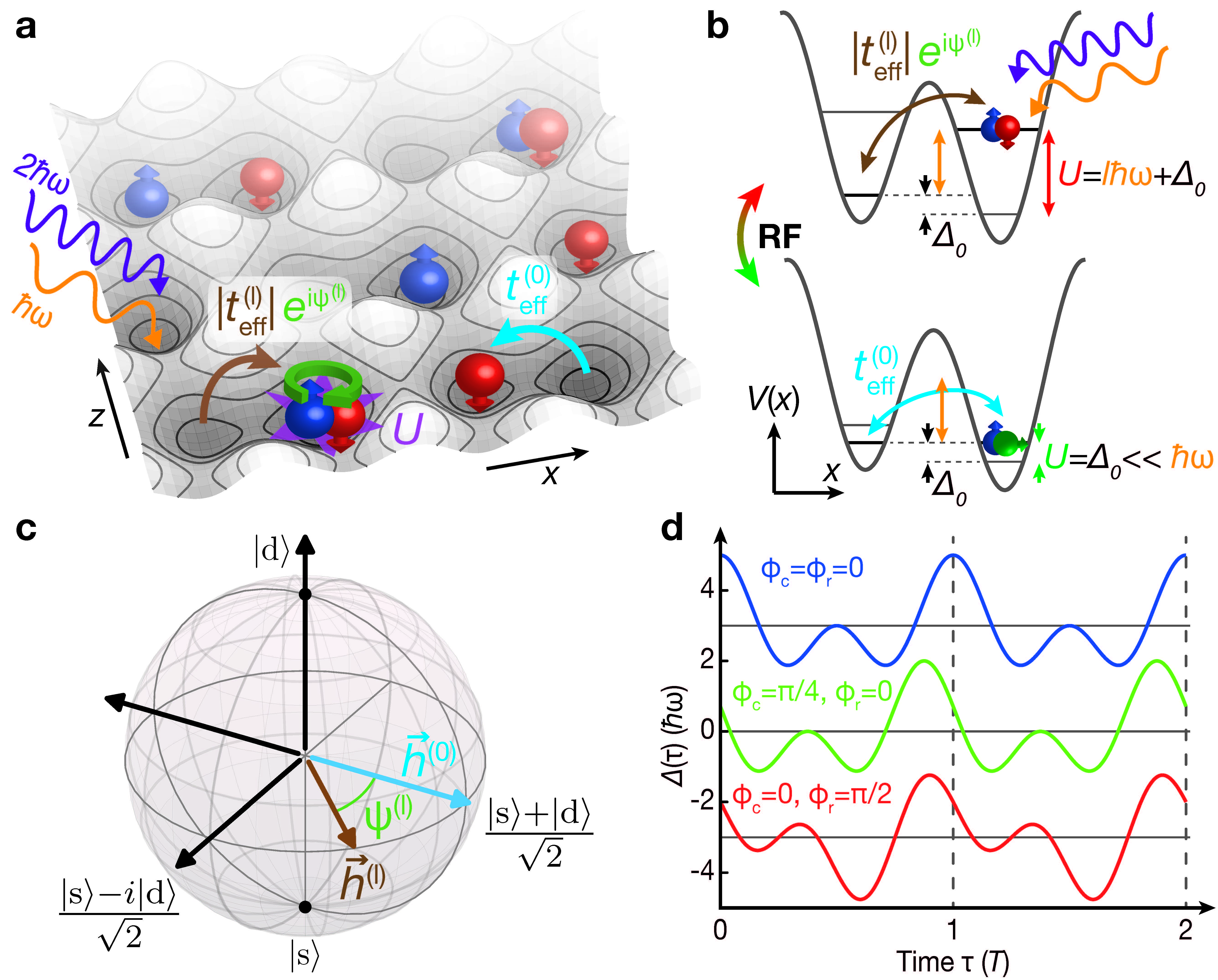}
	\caption{
	\textbf{Experimental setup and driving scheme.} 
	\textbf{a,} 
	Lattice potential in the $x$-$z$-plane consisting of individual dimers with an energy bias $\Delta_0$. The lattice position is sinusoidally modulated in the $x$-direction at two frequencies $\omega/(2\pi)$ and $2\omega/(2\pi)$ using a piezoelectric actuator (not shown). If the on-site interaction $U$ is tuned close to a resonance $U=l\hbar\omega+\Delta_0$, atoms pick up a phase $\psi^{(l)}$ in a density-assisted tunnelling process $t_{\text{eff}}^{(l)}$ compared to a single-particle hopping process $t_{\text{eff}}^{(0)}$ (with $\psi^{(0)}\approx 0$ for our parameters). 
	\textbf{b,}
	Schematic representation of the effective Hamiltonian in the double wells. The double occupancy state $\left|\text{d}\right\rangle=\left|0,\uparrow\downarrow\right\rangle$ is coupled to the singlet state $\left|\text{s}\right\rangle$ via a density-assisted tunnelling process induced by multi-photon processes of the resonant drive. Using a radio frequency (RF) pulse, it is possible to switch to a third internal state $\rightarrow$, for which the tunnel coupling is equivalent to the single-particle hopping amplitude $t_{\text{eff}}^{(0)}$.
	\textbf{c,} 
	Visualization of the effective two level system in \textbf{b} on a Bloch sphere. For $U=l\hbar\omega+\Delta_0$, the off-resonant Hamiltonian represented by the vector $\vec{h}^{(0)}$ is anti-aligned with the $x$-axis, while $\vec{h}^{(l)}$ is rotated around the $z$-axis by an angle $\psi^{(l)}$. 
	\textbf{d,} 
	Time-dependent energy offset $\Delta(\tau)$ between the two sites of the double well for the two-frequency driving scheme with $K_1=1$ and $K_2=0.5$. The common phase $\phi_{\text{c}}$ leads to a mere shift of the waveform in time, while the relative phase $\phi_{\text{r}}$ explicitly breaks time-reversal symmetry for $\phi_{\text{r}}\neq 0,\pi$. The waveforms are offset for clarity. 
	}
	\label{fig:1}
\end{figure}

In order to directly measure the matrix elements of the tunneling amplitude $\hat{t}_{\langle \textbf{i}, \textbf{j} \rangle,\bar{\sigma}}^{(l)}$ and the gauge field operator $\hat{\mathcal{A}}_{\langle \textbf{i}, \textbf{j} \rangle,\bar{\sigma}}^{(l)}$ in the experiment, we project the Hamiltonian (\ref{HeffDynGaugeFields}) onto a link $\langle \textbf{i}, \textbf{j}\rangle$ by realizing individual Hubbard dimers (Fig.\;\ref{fig:1}a). We introduce an asymmetry between the two sites of the dimer with a static energy bias $\Delta_0$, such that we can selectively address the density-assisted tunneling processes $t^{(l)}_{\text{eff,R}}$ and $t^{(l)}_{\text{eff,L}}$ with the drive. If $U$ is much larger than both $\Delta_0$ and the static tunneling $t$, the ground state for static double wells occupied by two atoms in states $\uparrow$ and $\downarrow$, respectively, is given by the singlet 
$\left|\text{s}\right\rangle=\left(\left|\uparrow,\downarrow\right\rangle - \left|\downarrow,\uparrow\right\rangle\right)/\sqrt{2}$. When driving the system resonantly such that $U\approx l\hbar\omega+\Delta_0$, the singlet is coupled to the double occupancy state $\left|\text{d}\right\rangle=\left|0,\uparrow\downarrow\right\rangle$ via $t^{(l)}_{\text{eff}}\equiv t^{(l)}_{\text{eff,R}}$ by absorbing $l$ photons from the drive (see Fig.\;\ref{fig:1}b and Supp.\;Fig.\;\ref{fig:S1}). The effective Hamiltonian in this two-level system can be written as
\begin{equation}
\hat{H}_{\text{eff}}^{(l)}=\vec{h}^{(l)}\cdot\vec{\sigma}+\frac{\delta^{(l)}}{2}\mathds{1}^{2\times 2}. 
\label{Heff}
\end{equation}
Here, $\vec{\sigma}=(\sigma_x,\sigma_y,\sigma_z)$ is the vector of the Pauli spin matrices and
\begin{equation}
\vec{h}^{(l)}=\left(-\sqrt{2}|t_{\text{eff}}^{(l)}|\cos(\psi^{(l)}),\sqrt{2}|t_{\text{eff}}^{(l)}|\sin(\psi^{(l)}),\delta^{(l)}/2\right), 
\end{equation}
where $\delta^{(l)}=U-l\hbar\omega-\Delta_0$ is the detuning from the $l$-th order resonance (for the measurement of $t^{(l)}_{\text{eff,L}}$ at $U\approx l\hbar\omega-\Delta_0$ see Supp.\;Fig.\;\ref{fig:S7}). 
In order to directly measure $\psi^{(l)}$ we perform an interference measurement, in which the single-particle phase $\psi^{(0)}$ acts as a reference (for our parameters $\psi^{(0)}\approx 0$, see Supp.\;Fig.\;\ref{fig:S3}). Instead of comparing $\psi^{(l)}$ to the phase acquired by single atoms in spatially separated dimers (see Fig.\;\ref{fig:1}a), we can conduct an interference measurement within each doubly occupied dimer by switching the state $\downarrow$ to a third internal state labeled $\rightarrow$ via a radio frequency (RF) pulse (see Fig.\;\ref{fig:1}b). The interaction between the spin states $\uparrow$ and $\rightarrow$ can be set to $U=\Delta_0\ll\hbar\omega$, such that the atoms experience the effective Hamiltonian in Eq.\;(\ref{Heff}) with $l=0$, which contains the single-particle tunnelling $t_{\text{eff}}^{(0)}$. On the Bloch sphere, the vector $\vec{h}^{(0)}$ representing the Hamiltonian $H_{\text{eff}}^{(0)}$ is pointing along the $x$-axis for $\delta^{(0)}=0$, while the resonant Hamiltonian $\vec{h}^{(l)}$ is rotated around the $z$-axis by an angle $\psi^{(l)}$ (see Fig.\;\ref{fig:1}c). To characterise any quantum state $\left|\varphi\right\rangle$, we can measure for both combinations of spins the fraction of double occupancies $\mathcal{D}=\left|\langle d\,|\varphi\rangle\right|^2$ and singlets $\mathcal{S}=\left|\langle s\,|\varphi\rangle\right|^2$. Here, $\left\langle ... \right\rangle$ denotes the average over the inhomogeneous distribution of $\Delta_0$ in different dimers resulting from the underlying harmonic trapping potential. 

In the experiment, we use a harmonically confined cloud of $N=41(4)\times 10^3$ ultracold fermionic $^{40}\text{K}$ atoms in a balanced mixture of two initial internal states $\uparrow$ and $\downarrow$, which are loaded into a dimerized, three-dimensional optical lattice (see Fig.\;\ref{fig:1}a). The sites constituting the dimers are connected with a static tunnelling amplitude of $t/h=260(30)\:\text{Hz}$ and are offset in energy by $\Delta_0/h=660(20)\:\text{Hz}$. Using a suitable loading procedure, $56(2)\%$ of the atoms occupy dimers that are populated by two opposite spins (see Methods). The interaction $U/h$ between atoms in states $\uparrow$ and $\downarrow$ can be tuned in a range between $5$ and $10\:\text{kHz}$ using a magnetic Feshbach resonance. The drive consists of a time-periodic modulation of the lattice position at two frequencies $\omega/(2\pi)=2.75\:\text{kHz}$ and $2\omega/(2\pi)=5.5\:\text{kHz}$, which in a co-moving frame correponds to a modulation of the energy offset $\Delta_{\text{tot}}(\tau)=\Delta_0+\Delta(\tau)$ within the dimers \cite{Desbuquois2017} with the time-dependent part
\begin{equation}
\Delta(\tau)=\hbar\omega K_1\cos(\omega\tau+\phi_{\text{c}})+2\hbar\omega K_2\cos(2\omega\tau+2\phi_{\text{c}}+\phi_{\text{r}}). 
\label{drive}
\end{equation}
Here, $K_1$ and $K_2$ are the dimensionless driving amplitudes and $\phi_{\text{c}}$ is a common phase which shifts the waveform in time without changing its shape (see Fig.\;\ref{fig:1}d). It can be set to zero by choosing an appropriate origin of time. In contrast, the relative phase $\phi_{\text{r}}$ explicitly breaks TRS for $\phi_{\text{r}}\neq 0,\pi$. Therefore, it will both affect the absolute value $|t_{\text{eff}}^{(l)}|$ and, crucially, lead to a non-trivial phase $\psi^{(l)}$ that cannot be eliminated by a suitable gauge choice. 

To derive the effective tunnelling matrix element for our driving scheme, we perform a high-frequency expansion in a rotating frame (see \cite{Bukov2015} and SI) and find 
\begin{equation}
t_{\text{eff}}^{(l)}=t e^{-il\phi_{\text{c}}}\sum_m \mathcal{J}_{-2m+l}(K_1)\mathcal{J}_m(K_2)e^{-im \phi_{\text{r}}}
\label{effTun}
\end{equation}
to lowest order, where $\mathcal{J}_m$ is the $m$-th order Bessel function. The effective tunnel coupling is given by the interference of all multi-photon processes in which $m$ photons are absorbed from the $2\omega$ drive and $2m-l$ photons are re-emitted into the $\omega$ drive, such that the total energy added to the system is $l\hbar\omega = U-\Delta_0$. In the experiment, we investigate the case $l=2$, for which the leading terms of the sum can be written as $t_{\text{eff}}^{(2)}=t(\alpha^{(2)}+\beta^{(2)} e^{-i\phi_{\text{r}}}$), where $\alpha^{(2)},\beta^{(2)}>0$ depend on $K_1$ and $K_2$ and we fixed the gauge such that $\phi_{\text{c}}=0$. It can be seen that if $\phi_{\text{r}}= 0$ or $\pi$ such that TRS is not broken, the tunnelling matrix element is real. Furthermore, if $\alpha^{(2)}=\beta^{(2)}$ and $\phi_{\text{r}}= \pi$, the tunnelling amplitude vanishes. Away from this singular point, $|t_{\text{eff}}^{(2)}|$ increases linearly with $\left|\alpha^{(2)}-\beta^{(2)}\right|$ and $\phi_{\text{r}}$ and is therefore forming a Dirac point in this generalized parameter space. At the same time, the Peierls phase $\psi^{(2)}$ has a vortex structure around the singularity. 

\begin{figure}[htb!]
	\includegraphics[width=89mm]{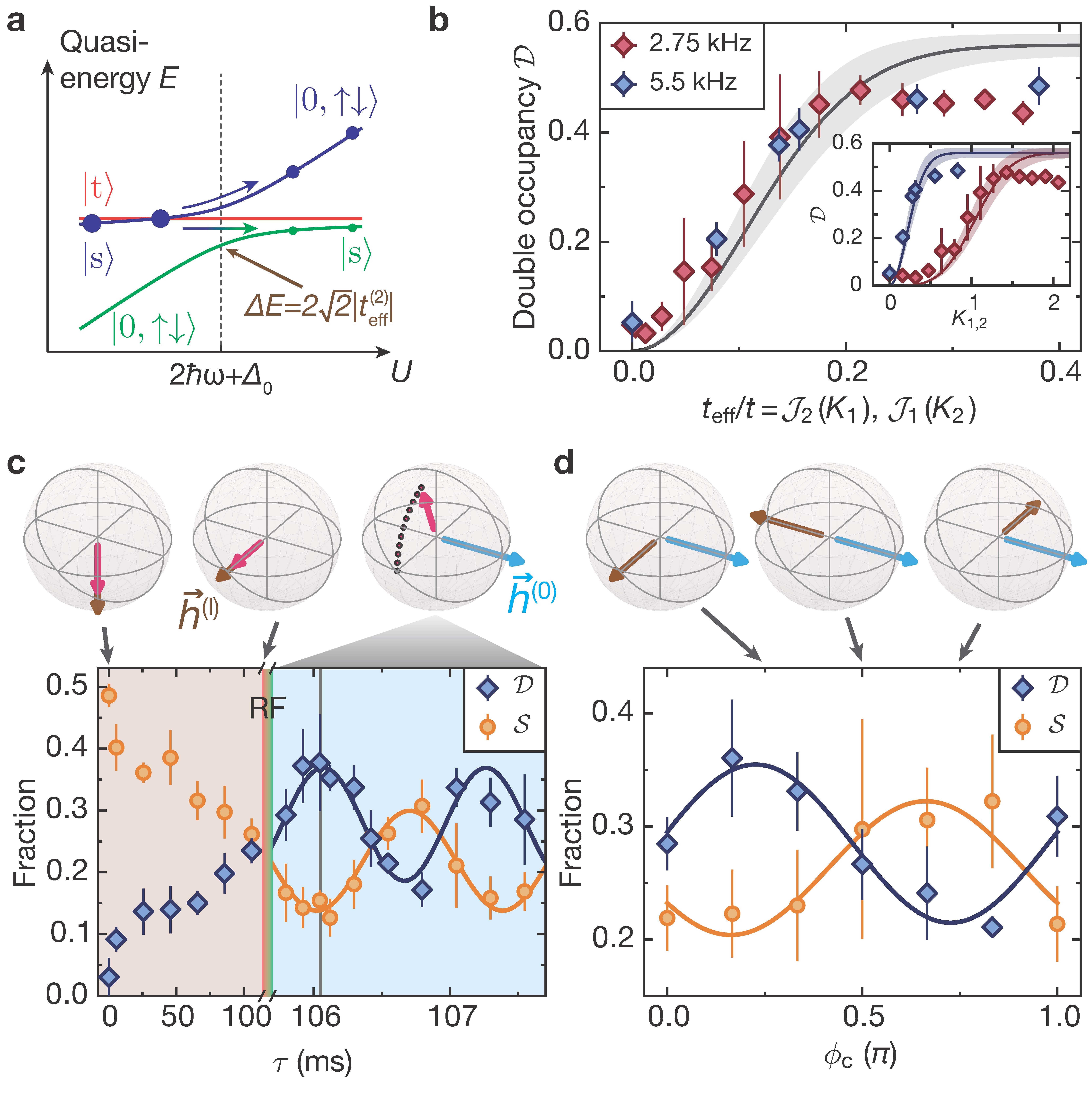}
	\caption{
	\textbf{Schemes to measure absolute value and phase of the effective tunnel coupling.} 
	\textbf{a,} 
	Quasi-energy spectrum around the resonance $U=2\hbar\omega+\Delta_0$, showing the avoided crossing between the singlet and double occupancy states with a gap given by $2\sqrt{2}|t_{\text{eff}}^{(2)}|$ on resonance. The absolute value $|t_{\text{eff}}^{(2)}|$ is measured by the amount of adiabatic transfer to the state $\left|0,\uparrow\downarrow\right\rangle$ in a Landau-Zener sweep of the interactions across the resonance. 
	\textbf{b,}
	Double occupancy fraction when ramping across the resonance for a single frequency drive at $\omega/(2\pi)=2.75\:\text{kHz}$ (red) or $2\omega/(2\pi)=5.5\:\text{kHz}$ (blue) as a function of the effective tunnelling $\mathcal{J}_2(K_1)$ or $\mathcal{J}_1(K_2)$, respectively. The line shows the theoretical transfer for a Landau-Zener sweep and the shaded area represents the uncertainty resulting from an imprecise knowledge of $\mathcal{D}_{\text{max}}$ and $\kappa$ (see main text). The inset depicts $\mathcal{D}$ as a function of $K_1$ or $K_2$, respectively.
	\textbf{c,} 
	Measurement scheme for the tunnelling phase $\psi_{\text{r}}^{(2)}$ illustrated for the case $K_1=0$, $K_2=0.79(1)$. The state (magenta arrow on the Bloch sphere) is first adiabatically following the Hamiltonian $H_{\text{eff}}^{(2)}$ and then quenched onto the Hamiltonian $H_{\text{eff}}^{(0)}$ using an RF pulse. Depending on the phase enclosed by the two Hamiltonians, the observables show coherent oscillations which have a relative phase shift of $\pi$. Solid lines are sinusoidal fits to the data. 
	\textbf{d,}
	Ramsey fringes as a function of the common phase when fixing the evolution time to the point where the Bloch vector rotated by an angle $\pi/2$ around $H_{\text{eff}}^{(0)}$ (grey line in \textbf{c}). Solid lines represent sinusoidal fits to the data with a fixed period of $\pi$. The tunnelling phase can be extracted as the phase of the Ramsey fringe, which gives $\psi_{\text{r}}^{(2)}=-0.05^{+0.04}_{-0.03}\pi$ and $\psi_{\text{r}}^{(2)}=0.83(8)\pi$ for the $\mathcal{D}$ and $\mathcal{S}$ fringes, respectively.
	Data points and error bars in \textbf{b} and \textbf{d} (\textbf{c}) denote mean and standard deviation of 5 (3) individual measurements. 
	}
	\label{fig:2}
\end{figure}

In our experiment, we measure both the absolute value of the effective tunnelling on the resonance $l=2$ (see Eq.\;(\ref{effTun})) and its phase compared to the single-particle tunnelling $l=0$. In order to quantify $|t_{\text{eff}}^{(2)}|$ without a bias resulting from the inhomogeneity of the harmonic trap, we perform a Landau-Zener type measurement. Starting from a singlet state in the static system at $U/h=5.41(7)\:\text{kHz}$, we first ramp up the modulation in $5.45\:\text{ms}$ while being detuned from the resonance and subsequently sweep the interactions over the avoided crossing to $U/h=7.9(1)\:\text{kHz}$ in $20\:\text{ms}$ (see Fig.\;\ref{fig:2}a). If the size of the gap at the resonance given by $2\sqrt{2}|t_{\text{eff}}^{(2)}|$ is large enough, we adiabatically follow the Floquet eigenstate and convert $\left|\text{s}\right\rangle$ to $\left|0,\uparrow\downarrow\right\rangle$ \cite{Desbuquois2017}. According to the Landau-Zener formula, the measured double occupancy fraction after the interaction sweep will be given by $\mathcal{D}=\mathcal{D}_{\text{max}}[1-\exp{(-\Gamma^2)}]$ with $\Gamma=|t_{\text{eff}}^{(2)}|/(\kappa\cdot t)$, where $\mathcal{D}_{\text{max}}=0.56(2)$ is the maximum value of $\mathcal{D}$ given by the initial preparation. The sensitivity of the measurement is characterised by $\kappa$, which is given by $\kappa=0.15(2)$ for our interaction ramp speed. To confirm the dependence of $\mathcal{D}$ on $|t_{\text{eff}}^{(2)}|$, we benchmark our gap measurement by driving only at a single frequency $\omega/(2\pi)$ or $2\omega/(2\pi)$. For $K_2=0$ or $K_1=0$, $|t_{\text{eff}}^{(2)}|$ reduces to $\mathcal{J}_2(K_1)$ or $\mathcal{J}_1(K_2)$, respectively. Fig.\;\ref{fig:2}b shows that the transfer fraction to the double occupancy state first increases with the magnitude of the effective tunnelling before it saturates for a gap size which corresponds to $|t_{\text{eff}}^{(2)}|\approx 0.2t$. This gives us a high sensitivity for small absolute values of the effective tunnelling. 

In order to directly measure the Peierls phase $\psi^{(2)}$, we implement a scheme similar to a Ramsey experiment. We design our protocol for the two-level system depicted in Fig.\;\ref{fig:1}c, in which we use the near- and off-resonant Hamiltonians represented by the vectors $\vec{h}^{(2)}$ and $\vec{h}^{(0)}$ as distinct rotation axes. Instead of scanning the evolution time of the state as in a typical Ramsey sequence, we vary the initial phase between $\vec{h}^{(2)}$ and $\vec{h}^{(0)}$ by changing the common phase $\phi_{\text{c}}$. The angle between the two rotation axes, which determines the interference fringes for the populations of $\left|\text{d}\right\rangle$ and $\left|\text{s}\right\rangle$, is given by $\psi^{(2)}=-2\phi_{\text{c}}+\psi_{\text{r}}^{(2)}(K_1,K_2,\phi_{\text{r}})$. In addition to $\phi_{\text{c}}$, it contains the non-trivial part of the Peierls phase $\psi_{\text{r}}^{(2)}$ given by the amplitudes and relative phase of the two-frequency modulation (see Eq.\;(\ref{effTun})). By varying $\phi_{\text{c}}$, we can extract the Peierls phase $\psi_{\text{r}}^{(2)}$ from the phase of the resulting fringes. 

More precisely we first prepare an eigenstate of the Hamiltonian $H_{\text{eff}}^{(2)}$ (see Eq.\;(\ref{Heff})) for $\delta^{(2)}=0$, which is given by $(-e^{i\psi^{(2)}},1)/\sqrt{2}$. This is achieved by ramping up the drive within $5.45\:\text{ms}$ away from the resonance followed by a sweep of the interactions on resonance to $U/h=6.23(8)\:\text{kHz}$ within $100\:\text{ms}$ (see Fig.\;\ref{fig:2}c). After that, we project the system onto the off-resonant Hamiltonian $H_{\text{eff}}^{(0)}$ at $\delta^{(0)}=0$. The quench is achieved by applying an RF pulse which lasts $9.5\;\mu\text{s}$ and converts $95(4)\%$ of the $\downarrow$ atoms to spin $\rightarrow$. If $H_{\text{eff}}^{(0)}$ is not (anti-)parallel to $H_{\text{eff}}^{(2)}$, the state will start to rotate around the new Hamiltonian, leading to oscillations of the singlet and double occupancy fractions (see Fig.\;\ref{fig:2}c). When fixing the evolution time to the point where the Bloch vector has rotated by an angle of $\pi/2$ around $H_{\text{eff}}^{(0)}$, we observe Ramsey fringes for the observables as a function of $\phi_{\text{c}}$ given by $\mathcal{D}(\phi_{\text{c}}),\mathcal{S}(\phi_{\text{c}})=\left[1\pm \sin{\left(-2\phi_{\text{c}}+\psi_{\text{r}}^{(2)}\right)}\right]/2$ from which we can directly extract $\psi_{\text{r}}^{(2)}$ (see Fig.\;\ref{fig:2}d). Due to the evolution of the state during the initial preparation of the eigenstate of $H_{\text{eff}}^{(2)}$, we measure an overall phase offset of the Ramsey fringes, which we determine in independent measurements to be $-0.15(4)\pi$ (see Methods). For all data shown in the following, the tunnelling phase was extracted from the Ramsey fringes for $\mathcal{D}$. 

\begin{figure}[htb]
	\includegraphics[width=89mm]{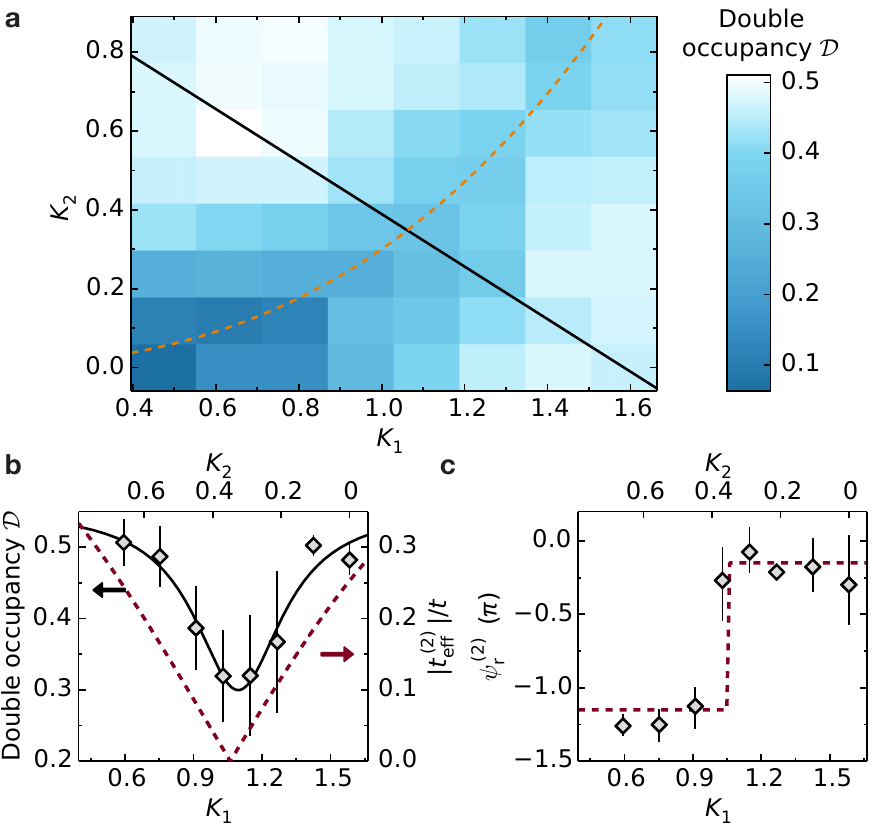}
	\caption{
	\textbf{Gap closing in the $K_1$-$K_2$ parameter space.} 
	\textbf{a,} 
	Double occupancy fraction when ramping across the $U=2\hbar\omega+\Delta_0$ resonance as a function of the modulation amplitudes $K_1$ and $K_2$ for $\phi_{\text{r}}=1.00(1)\pi$. 
	The orange dashed line marks the theoretical value for which $|t_{\text{eff}}^{(2)}|=0$ according to Eq.\;(\ref{effTun}). 
	\textbf{b,}
	Cut along the black line indicated in \textbf{a}. The solid line is a Lorentzian fit to the data (left axis), from which we extract the amplitude at the minimum gap to be $K_1=1.10^{+0.07}_{-0.11}$. The red dashed line shows the theoretical value of $|t_{\text{eff}}^{(2)}|$ (right axis), which depends linearly on the driving amplitude around the gap closing. 
	\textbf{c,} 
	Phase $\psi_{\mathrm{r}}^{(2)}$ of the effective tunnelling along the same cut as in \textbf{b}, showing a jump of $\pi$ between $K_1=0.91(1)$ and $1.03(1)$. The according sign change of $t_{\text{eff}}^{(2)}$ demonstrates the true gap closing in \textbf{b}. The red dashed line is the theoretical value of $\psi_{\mathrm{r}}^{(2)}$ according to Eq.\;(\ref{effTun}), taking into account the experimental phase offset of $-0.15(4)\pi$.   
	Data points and error bars in \textbf{a} and \textbf{b} denote mean and standard deviation of 5 individual measurements. Mean values in \textbf{c} are derived from a sinusoidal fit to the Ramsey fringes and errors denote the standard deviation obtained from a resampling method (see Methods). 
	}
	\label{fig:3}
\end{figure}

We begin our investigation of the effective tunnel coupling induced by the two-frequency drive by mapping out the transition for which $|t_{\text{eff}}^{(2)}|=0$ in the $K_1$-$K_2$ parameter space. As discussed above, this occurs at the TR-symmetric point $\phi_{\text{r}}=\pi$ and to lowest order for $\alpha^{(2)}=\beta^{(2)}$, i.e. $\mathcal{J}_2(K_1)\mathcal{J}_0(K_2)=\mathcal{J}_0(K_1)\mathcal{J}_1(K_2)$. Fig.\;\ref{fig:3}a shows the result of the gap measurement in the $K_1$-$K_2$ parameter space following the experimental protocol in Fig.\;\ref{fig:2}a,b. The gap shows a clear minimum along the diagonal, separating two distinct regions with large values of $|t_{\text{eff}}^{(2)}|$. The gap closing nicely follows the theoretical prediction derived from Eq.\;(\ref{effTun}) without free parameters (see also Supp.\;Fig.\;\ref{fig:S2}). While the double occupancy fraction goes to almost zero for small values of both $K_1$ and $K_2$, the minimum is less pronounced if both amplitudes are high. In this region, the two-level approximation in Eq.\;(\ref{Heff}) breaks down and the singlet is transformed into the other double occupancy state $\left|\uparrow\downarrow,0\right\rangle$ during the interaction sweep. This is demonstrated by a full numerical simulation of the gap measurement protocol (see Supp.\;Fig.\;\ref{fig:S5}) and results from the proximity of the resonances at $U=l\hbar\omega\pm\Delta_0$, which can be avoided by choosing higher modulation frequencies. 

The line along which the gap closes separates two distinct regions in parameter space, which are characterized by a $\mathbb{Z}_2$-invariant (see SI). When going from top left to bottom right in Fig.\;\ref{fig:3}a, the parameter $\alpha^{(2)}-\beta^{(2)}$ and hence the tunneling matrix element $t_{\text{eff}}^{(2)}$ change from negative to positive values. At the phase transition, $\alpha^{(2)}-\beta^{(2)}=0$ such that $|t_{\text{eff}}^{(2)}|=0$ (see Fig.\;\ref{fig:3}b showing a cut along the black line indicated in Fig.\;\ref{fig:3}a). The sign change of the effective tunnelling amplitude can be demonstrated by measuring its phase across the transition line, which exhibits a sharp jump by $\pi$ for the critical values of $K_1$ and $K_2$ where the gap closes (see Fig.\;\ref{fig:3}c). This in turn proves that the gap fully closes, since the tunnelling amplitude is continuous in the modulation parameters. 

After mapping out the gap closing in the parameter space of the driving amplitudes, we additionally investigate the influence of the relative modulation phase $\phi_{\text{r}}$. To this end, we always fix the parametrization in the $K_1$-$K_2$ space to be along the black line in Fig.\;\ref{fig:3}a. If we expand the tunnel coupling around the point where the gap closes up to linear order in $\tilde{K}_1=K_1-K_{1,\text{crit}}$ and $\tilde{\phi}_{\text{r}}=\phi_{\text{r}}-\pi$, we find $t_{\text{eff}}^{(2)}=t[c_K \tilde{K}_1+i c_{\phi}\tilde{\phi}_{\text{r}}]/\sqrt{2}$ (see SI). Here, the numerical factors $K_{1,\text{crit}}=1.06(1)$, $c_K=0.537(1)$ and $c_{\phi}=0.123(1)$ depend on the $K_1$-$K_2$ parametrization. For $\delta^{(2)}=0$, the low-energy Hamiltonian around the gap closing point can therefore be written as
\begin{equation}
H_{\text{Dirac}}=-t c_K \tilde{K}_1 \sigma_x+t c_{\phi} \tilde{\phi}_{\text{r}} \sigma_y. 
\label{Hdirac}
\end{equation} 
This is a Dirac Hamiltonian in the driving parameters, which only affects the density-assisted tunnelling processes, while the single-particle hopping remains trivial. 

\begin{figure*}
	\includegraphics[width=178mm]{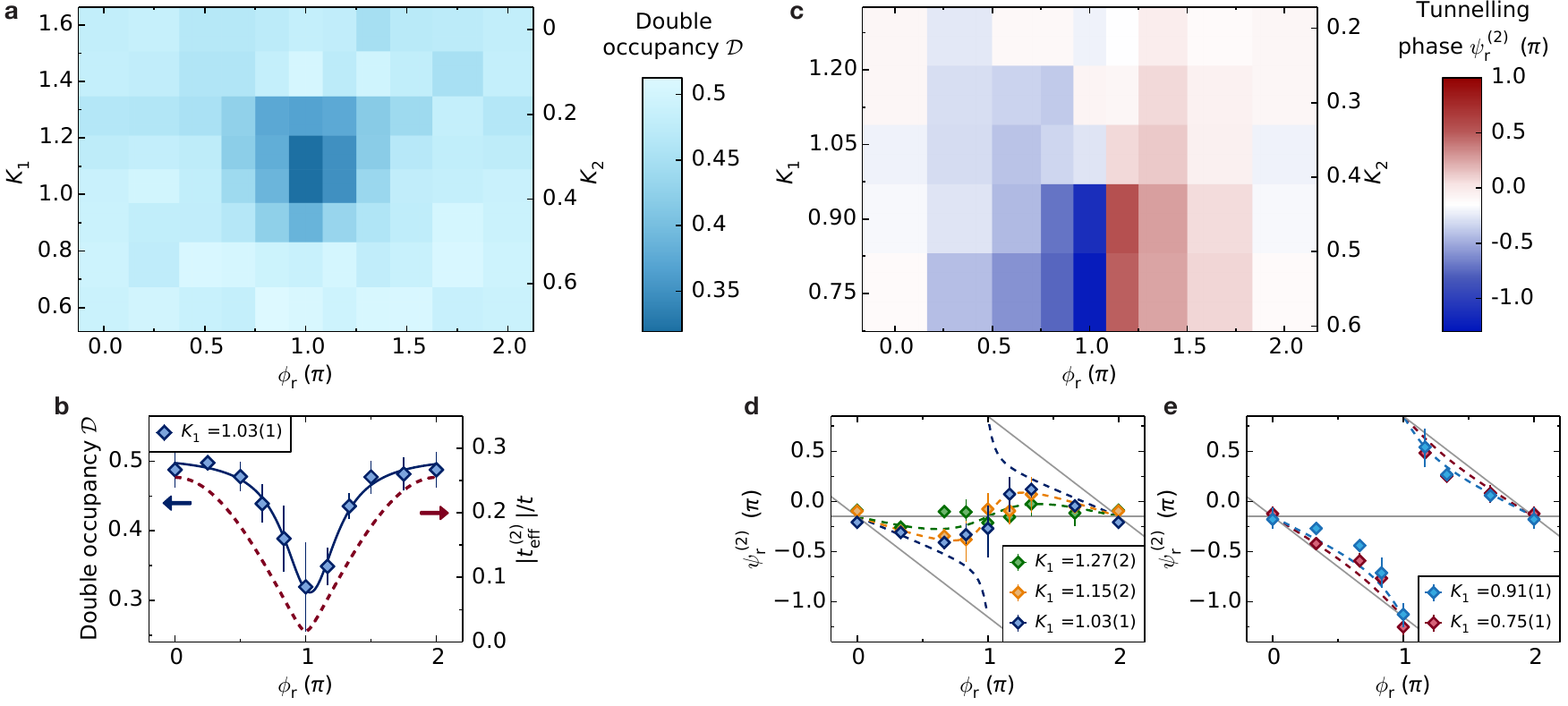}
	\caption{
	\textbf{Dirac point and Peierls phase vortex.} 
	\textbf{a,} 
	Double occupancy fraction when ramping across the $U=2\hbar\omega+\Delta_0$ resonance as a function of the relative phase $\phi_{\text{r}}$ and the driving amplitudes $K_1$-$K_2$. The amplitudes are parametrized along the black line in Fig.\;\ref{fig:3}a. The gap is closing at a singular point in parameter space. 
	\textbf{b,}
	Cut through \textbf{a} at $K_1=1.03(1)$. The solid line is a Lorentzian fit to the data (left axis), from which we extract the relative phase where the gap closes to be $\phi_{\text{r}}=1.03(6)\pi$. The red dashed line shows the theoretical value of the gap (right axis), which depends linearly on the relative phase around the gap closing. 
	\textbf{c,} 
	Tomography of $\psi_{\mathrm{r}}^{(2)}$ as a function of the relative phase $\phi_{\text{r}}$ and the driving amplitudes $K_1$-$K_2$, showing a vortex structure around the Dirac point. 
	\textbf{d,e,} 
	Cuts through \textbf{c} for fixed values of $K_1$ (see legends). Dashed lines show the theoretical value of $\psi_{\mathrm{r}}^{(2)}$ according to Eq.\;(\ref{effTun}), taking into account the experimental phase offset of $-0.15(4)\pi$ (gray horizontal line). 
	Data points and error bars in \textbf{a} and \textbf{b} denote mean and standard deviation of 5 individual measurements. Mean values in \textbf{c}, \textbf{d} and \textbf{e} are derived from a sinusoidal fit to the Ramsey fringes and errors denote the standard deviation obtained from a resampling method (see Methods). 
	Data at $\phi_{\text{r}}=2\pi$ is duplicated from $\phi_{\text{r}}=0$. 
	}
	\label{fig:4}
\end{figure*}

Fig.\;\ref{fig:4}a shows the measurement of the gap near the Dirac point located at $K_{1,\text{crit}}$ and $\phi_{\text{r}}=\pi$. It demonstrates that $|t_{\text{eff}}^{(2)}|$ has a clear minimum at the singularity and increases away from it. The gap closes at $\phi_{\text{r}}=1.03(6)\pi$ as expected from theory (see analytical results in Fig.\;\ref{fig:4}b and Supp.\;Fig.\;\ref{fig:S4} and a numerical simulation of the gap measurement in Supp.\;Fig.\;\ref{fig:S6}). In addition, Fig.\;\ref{fig:4}c shows a full tomography of the tunnelling phase $\psi_{\text{r}}^{(2)}$ around the Dirac point. It has a vortex structure and the phase increases by $2\pi$ when going clockwise around the singularity. For high values of $K_1>K_{1,\text{crit}}$, $\psi_{\text{r}}^{(2)}$ only changes little as a function of the relative phase (see Fig.\;\ref{fig:4}d). In this regime, the driving component at $\omega$ is dominant, which corresponds to the lower right corner in Fig.\;\ref{fig:3}a. Around the critical point $K_1=K_{1,\text{crit}}$ and $\phi_{\text{r}}=\pi$, the tunnelling phase is very sensitive to the exact driving parameters and suddenly jumps from $0$ to $\pi$ when lowering $K_1$ (see also Fig.\;\ref{fig:3}c). For $K_1<K_{1,\text{crit}}$, we enter the upper left region in Fig.\;\ref{fig:3}a and suddenly observe a running phase $\psi_{\text{r}}^{(2)}$, which means that the state vector is winding once around the Bloch sphere when $\phi_{\text{r}}$ is swept from $0$ to $2\pi$ (see Fig.\;\ref{fig:4}e). While the detailed shape of the phase vortex depends on the parametrization of the driving waveform in Eq.\;(\ref{drive}), the phase difference between two configurations at the TR-symmetric points is forced to be either $0$ or $\pi$. This quantity is therefore a $\mathbb{Z}_2$-invariant which can be used to characterise the corresponding regimes (see Fig.\;\ref{fig:3}a and SI). 

In future experiments, the full control over both the amplitude and Peierls phase of the density-dependent tunnelling matrix element demonstrated in this work can be further extended by introducing a temporal or spatial dependence for the driving parameters, which maps the Dirac point into another parameter space. 
In addition, it is straightforward to couple the individual dimers by allowing for tunnelling in all directions in order to study the intriguing interplay between the interaction induced gauge field and the atomic density. 
Recent experiments have shown that driven many-body systems can be well understood in the effective Hamiltonian picture and that problems associated with interacting Floquet systems such as heating can be mitigated in certain lattice geometries \cite{Gorg2018,Messer2018,Sandholzer2018}. 
In higher dimensions, a variety of phenomena related to density-dependent gauge fields could be studied, such as anyonic statistics in one dimension \cite{Keilmann2011,Greschner2015,Cardarelli2016,Strater2016} or flux attachment \cite{Barbiero2018}. 
Finally, the symmetry between atoms in different internal states can be broken by using a spin-selective drive \cite{Jotzu2015}, such that distinct gauge and matter particles can be identified. As shown in the SI, using such a driving scheme involving two modulation frequencies enables to engineer both a dynamical $\mathbb{Z}_2$ gauge field $\hat{\mathcal{A}}_{\langle \textbf{i}, \textbf{j} \rangle}=\text{diag}\,(0,\pi)$, which only requires real-valued tunneling matrix elements \cite{Barbiero2018,Schweizer2019}, as well as a $\mathbb{Z}_3$ gauge field  $\hat{\mathcal{A}}_{\langle \textbf{i}, \textbf{j} \rangle}=\text{diag}\,(0, 2\pi/3, 4\pi/3)$.


\vspace{5mm}

\textbf{Acknowledgements} 
We thank L. Barbiero, A. Bermudez, A. Eckardt, N. Goldman, F. Grusdt, Y. Murakami, M. Rizzi, L. Santos, K. Viebahn, P. Werner and O. Zilberberg for insightful discussions and K. Viebahn and O. Zilberberg for a careful reading of the manuscript. 
We acknowledge SNF (project numbers 169320 and 182650), NCCR-QSIT, QUIC (Swiss State Secretary for Education, Research and Innovation contract number 15.0019) and ERC advanced grant TransQ (Project Number 742579) for funding. 

\textbf{Author Contributions} 
The data were measured and analysed by F.G., K.S. and J.M. The theoretical framework and measurement scheme were developed by F.G. All work was supervised by T.E. All authors contributed to planning the experiment, discussions and the preparation of the manuscript. 

\textbf{Author Information} 
The authors declare no competing financial interests. Correspondence and requests for materials should be addressed to T.E. (esslinger@phys.ethz.ch).

\clearpage

\makeatletter

\setcounter{section}{0}
\setcounter{subsection}{0}
\setcounter{figure}{0}
\setcounter{equation}{0}
\setcounter{table}{0}
\setcounter{NAT@ctr}{0}

\renewcommand{\theequation}{M\@arabic\c@equation}
\renewcommand{\theHequation}{MH\theequation}

\makeatother

\section*{METHODS} 

\subsection{Driving scheme and effective Hamiltonian}
In the following, we derive the effective Hamiltonian for a Fermi-Hubbard model, which is driven at two frequencies that are resonant with the onsite interaction $U$. We show that this scheme is suited to control both the amplitude and Peierls phase of density-assisted tunneling processes independently from the single-particle hopping. For simplicity, we discuss the case of a one-dimensional lattice. For a more details and extensions of the scheme to realize dynamical $\mathbb{Z}_2$ and $\mathbb{Z}_3$ gauge fields see SI. 

The full time-dependent Hamiltonian can be written as $\hat{H}(\tau)=\hat{H}_0+\hat{V}(\tau)$, where the static part corresponds to the usual Fermi-Hubbard model
\begin{equation}
\hat{H}_0=-t \sum_{j,\sigma} (\hat{c}^{\dagger}_{j+1,\sigma}\hat{c}_{j\sigma}+\text{h.c.})+U \sum_j \hat{n}_{j\uparrow}\hat{n}_{j\downarrow}
\label{staticFHMMethods}
\end{equation}
and the drive is given by
\begin{equation}
\hat{V}(\tau)=-f(\tau)\sum_{j,\sigma} j \hat{n}_{j\sigma}
\label{DriveOpManyBodyMethods}
\end{equation}
in a frame that is co-moving with the shaken lattice. For the two-frequency modulation scheme, the oscillating inertial force reads
\begin{eqnarray}
f(\tau)&=&\hbar\omega K_1\cos(\omega\tau+\phi_{\text{c}})\notag\\
			 &+& 2\hbar\omega K_2\cos(2\omega\tau+2\phi_{\text{c}}+\phi_{\text{r}})
\label{DriveTwoFreqManyBodyMethods}
\end{eqnarray}
in analogy to the site offset in the case of a double well (see Eq.\;(\ref{drive}) of the main text). Using Floquet theory, we can derive an effective static Hamiltonian around the resonance $U=l\hbar\omega$ ($l\in\mathbb{Z}$) \cite{Bukov2015}, which yields to lowest order
\begin{eqnarray}
\hat{H}^{(l)}_{\text{eff}}=&-&\sum_{j,\sigma} \left[\left(t_{\text{eff}}^{(0)}\hat{a}_{\langle j,j+1 \rangle,\bar{\sigma}}+ t_{\text{eff},\text{R}}^{(l)}\hat{b}_{\langle j,j+1 \rangle,\bar{\sigma}\text{R}} \right. \right. \notag\\
&+& \left. \left. \left[t_{\text{eff},\text{L}}^{(l)}\right]^* \hat{b}_{\langle j,j+1 \rangle,\bar{\sigma}\text{L}}\right) \hat{c}^{\dagger}_{j+1,\sigma}\hat{c}_{j\sigma}+\text{h.c.}\right] \notag\\
&+& (U-l\hbar\omega) \sum_j \hat{n}_{j\uparrow}\hat{n}_{j\downarrow}. 
\label{HeffManyMethods}
\end{eqnarray}
Here, single-particle tunnelings associated with the operator $\hat{a}_{\langle i,j \rangle,\sigma} = (1-\hat{n}_{i\sigma})(1-\hat{n}_{j\sigma})+\hat{n}_{i\sigma}\hat{n}_{j\sigma}$ are described by the matrix element $t_{\text{eff}}^{(0)}$, while density-assisted processes that create or annihilate a double occupancy via hopping to the right (R) or left (L) corresponding to the operators $\hat{b}_{\langle i,j \rangle,\sigma\text{R}} = (1-\hat{n}_{i\sigma})\hat{n}_{j\sigma}$ and $\hat{b}_{\langle i,j \rangle,\sigma\text{L}} = \hat{n}_{i\sigma}(1-\hat{n}_{j\sigma})$, respectively, are governed by $t_{\text{eff},\text{R/L}}^{(l)}$. The matrix elements are given by
\begin{equation}
t_{\text{eff},\text{R/L}}^{(l)} = te^{-il\phi_{\text{c}}}\sum_m \mathcal{J}_{-2m\pm l}(K_1)\mathcal{J}_m(K_2)e^{\mp im \phi_{\text{r}}}
\label{teffMethods}
\end{equation}
for $l\in\mathbb{Z}$ (see Eq.(\ref{effTun}) of the main text).

For $U=l\hbar\omega$, the effective Hamiltonian in Eq.\;(\ref{HeffManyMethods}) can be written as in Eq.\;(\ref{HeffDynGaugeFields}) in the main text
\begin{equation}
\hat{H}^{(l)}_{\text{eff}}=-\sum_{j,\sigma} \left(\hat{t}_{\langle j,j+1 \rangle,\bar{\sigma}}^{(l)} e^{i\hat{\mathcal{A}}_{\langle j,j+1 \rangle,\bar{\sigma}}^{(l)}} \hat{c}^{\dagger}_{j+1,\sigma}\hat{c}_{j\sigma}+\text{h.c.}\right),
\label{HeffManySpinMethods}
\end{equation}
where the tunneling amplitudes and phases are described by the operators
\begin{eqnarray}
\hat{t}_{\langle i,j \rangle,\sigma}^{(l)} &=& |t_{\text{eff}}^{(0)}|\:\hat{a}_{\langle i,j \rangle,\sigma} + |t_{\text{eff},\text{R}}^{(l)}|\:\hat{b}_{\langle i,j \rangle,\sigma\text{R}} \notag\\ 
&+& |t_{\text{eff},\text{L}}^{(l)}|\:\hat{b}_{\langle i,j \rangle,\sigma\text{L}} \label{TunAmpOp}\\
\hat{\mathcal{A}}_{\langle i,j \rangle,\sigma}^{(l)} &=& \psi^{(0)}\:\hat{a}_{\langle i,j \rangle,\sigma} + \psi_{\text{R}}^{(l)}\:\hat{b}_{\langle i,j \rangle,\sigma\text{R}} \notag\\
&-& \psi_{\text{L}}^{(l)}\:\hat{b}_{\langle i,j \rangle,\sigma\text{L}}. \label{GaugeFieldOp}
\end{eqnarray}
In Eq.\;(\ref{HeffManySpinMethods}) it becomes apparent that atoms in the opposite spin state $\bar{\sigma}$ act as link variables and determine both the hopping amplitude and phase for atoms in state $\sigma$. Since there are three distinct tunneling processes corresponding to the operators $\hat{b}_{\langle i,j \rangle,\sigma\text{L}}$, $\hat{a}_{\langle i,j \rangle,\sigma}$ and $\hat{b}_{\langle i,j \rangle,\sigma\text{R}}$ for which the occupation difference between sites $i$ and $j$ equals $-1$, $0$ or $1$, respectively, it is natural to express the operators in Eqs.\;(\ref{TunAmpOp}) and (\ref{GaugeFieldOp}) in the eigenbasis of the spin-1 operator $\hat{\tau}_{\langle i,j \rangle,\sigma}^z = \hat{n}_{i\sigma}-\hat{n}_{j\sigma}$, which has eigenvalues $m_{\langle i,j \rangle,\sigma}\in \{-1,0,1\}$. In this basis, the tunneling and gauge field operators are given by $\hat{t}_{\langle i,j \rangle,\sigma}^{(l)}=\text{diag}(|t_{\text{eff},\text{L}}^{(l)}|,|t_{\text{eff}}^{(0)}|,|t_{\text{eff},\text{R}}^{(l)}|)$ and $\hat{\mathcal{A}}_{\langle i,j \rangle,\sigma}^{(l)}=\text{diag}(-\psi_{\text{L}}^{(l)},\psi^{(0)},\psi_{\text{R}}^{(l)})$, respectively. All tunneling amplitudes and phases can be tuned independently via the driving amplitudes $K_1$ and $K_2$ and the relative phase $\phi_{\text{r}}$. 

\subsection{Optical lattice}
We perform our experiments in a three-dimensional, optical lattice which is formed by a combination of four orthogonal, retro-reflected laser beams at a wavelength of $\lambda=1064\:\text{nm}$ (for more details, see earlier work \cite{Tarruell2012}). While the beams $\overline{\text{X}}$ and Y are effectively not interfering with any other beam due to a frequency detuning, the beams X and Z are interfering with each other and are actively phase-stabilized to $\varphi=0.00(3)\pi$. The resulting potential for the atoms is given by
\begin{eqnarray} V(x,y,z) & = & -V_{\overline{X}}\cos^2(k
x+\theta/2)-V_{X} \cos^2(k
x)\nonumber\\
&&-V_{Y} \cos^2(k y) -V_{Z} \cos^2(k z) \label{Lattice} \\
&&-2\alpha \sqrt{V_{X}V_{Z}}\cos(k x)\cos(kz)\cos\varphi \nonumber
\end{eqnarray}
where $k=2\pi/\lambda$ and $V_{\overline{X},X,Y,Z}$ are the lattice depths in units of the recoil energy $E_{\text{R}}=h^2/2m\lambda^2$ of each laser beam in the three different directions $x,y,z$  ($h$ is the Planck constant and $m$ the mass of the atoms). Both the interference term $\alpha=1.01(1)$ and the individual lattice depths $V_{\overline{X},X,Y,Z}$ are calibrated via amplitude modulation of the lattice depth with a $^{87}\text{Rb}$ Bose-Einstein condensate. To account for systematic errors from the calibration and fluctuations of the lattice depths, we include a relative error of $4.9\%$ (for $\overline{\text{X}}$ and X) or $2.3\%$ (for Y and Z) on the lattice depths for the calculation of the tight-binding parameters. In addition, we take into account a relative uncertainty of the magnetic field of $10^{-4}$. In the final lattice configuration, the depths are given by $V_{\overline{X},X,Y,Z}=[22(1),1.00(5),38.8(9),29.3(7)]\;E_{\text{R}}$. The corresponding potential consists of an array of individual double wells aligned in the $x$-direction with an intra-dimer tunnel coupling $t/h=260(30)\:\text{Hz}$. All dynamics between different dimers are suppressed by adjusting the inter-dimer tunnelling amplitudes in all spatial directions to be below $t/h=2\:\text{Hz}$. In addition, the phase $\theta$ in Eq.\;(\ref{Lattice}) is adjusted to $1.0050(1) \pi$, which introduces a static energy bias between the two sites of the double well of $\Delta_0/h=660(20)\:\text{Hz}$. Due to the underlying harmonic confinement with trapping frequencies of $\omega_{x,y,z}/(2\pi)=[121(2),107(1),151(2)]\:\text{Hz}$, an additional inhomogeneous site offset is introduced in the dimers. For a double well which is located at a typical distance of 20 lattice sites away from the center of the trap, the additional tilt is on the order of $\Delta_0/2$. 

\subsection{Preparation of the atoms in the double wells}
The preparation procedure of two distinguishable Fermions in the double wells is very similar to earlier work\;\cite{Desbuquois2017}. In brief, the starting point of our experiment is a balanced mixture of atoms in the $F=9/2, m_F=-9/2$ and $F=9/2,m_F=-7/2$ hyperfine states of $^{40}\text{K}$ (called $\uparrow$ and $\rightarrow$ in the main text), which are confined in an optical harmonic trap. After evaporatively cooling the atoms, we end up with $N=41(4)\times 10^3$ atoms at a temperature $T/T_{\text{F}}=0.09(2)$ ($T_{\text{F}}$ denotes the Fermi-temperature). After this, we tune the s-wave scattering length between the atoms to be very strongly attractive $a\rightarrow -\infty$ by means of a magnetic Feshbach resonance located at $202.1\:\text{G}$. Then, we first load the atoms into a checkerboard lattice with $V_{\overline{X},X,Y,Z}=[0,3,7,3]\;E_{\text{R}}$ within $200\:\text{ms}$ followed by a second ramp to a deep checkerboard lattice with $V_{\overline{X},X,Y,Z}=[0,30,40,30]\;E_{\text{R}}$ in $20\:\text{ms}$. Due to the strong attractive interactions, $56(2)\%$ of the atoms form a doubly occupied site, while no site is occupied by more than two atoms due to Pauli-blocking. The next step is to perform a Landau-Zener sweep with a radio frequency (RF) pulse to transfer the atoms in the $F=9/2,m_F=-7/2$ state to the $F=9/2,m_F=-5/2$ state (called $\downarrow$ in the main text). The interactions of the mixture in the $m_F=-9/2$ and $m_F=-5/2$ states can be tuned by a second magnetic Feshbach resonance at $224.2\:\text{G}$, which allows us to access strong repulsive interactions with $a>175\;a_0$ ($a_0$ is the Bohr radius). We adjust the scattering length to $a=237(1)\;a_0$ and subsequently split the wells of the checkerboard lattice into two sites by ramping to the final lattice configuration with $V_{\overline{X},X,Y,Z}=[22(1),1.00(5),38.8(9),29.3(7)]\;E_{\text{R}}$ in $20\:\text{ms}$. Here, the atoms interact with an on-site interaction energy $U/h=5.41(7)\:\text{kHz}$ and this is the starting point of the experiments. 

\subsection{Periodic driving}
The driving is implemented as in previous work \cite{Desbuquois2017, Gorg2018}. We sinusoidally modulate the position of the retro-reflecting mirror with a piezoelectric actuator along the direction of the dimers. As a result, the entire lattice potential is moving in space and the time-dependent potential for the two-frequency drive is given by $V(x,y,z,\tau)\equiv V(x-A_1\cos(\omega\tau+\phi_{\text{c}})-A_2\cos(2\omega\tau+2\phi_{\text{c}}+\phi_{\text{r}}),y,z)$. Here, $A_n$ are the real-space amplitudes of the modulation at frequency $n\omega/(2\pi)=n\cdot 2.75\:\text{kHz}$ ($n=1,2$). They are related to the normalised drive amplitude by $K_n=m\,A_n\,\omega\,d/\hbar$, where $d$ is the distance between the two sites of the double well. In our lattice geometry, $d\neq \lambda/2$ and the distance has to be calculated for the specific lattice geometry used in the experiment. For that, we calculate the maximally localised Wannier functions as eigenstates of the band-projected position operator and extract their center-of-mass position. For our lattice configuration $V_{\overline{X},X,Y,Z}=[22(1),1.00(5),38.8(9),29.3(7)]\;E_{\text{R}}$ we find $d=0.792(7)\cdot\lambda/2$. Furthermore, $\phi_{\text{c}}$ and $\phi_{\text{r}}$ are the common and relative phase of the two-frequency drive. During the modulation, we make sure that the phase relation between the X- and Z-beams is stabilized to $\varphi=0.00(3)\pi$ by modulating the frequency of the respective incoming laser beams using acousto-optical modulators. This phase modulation of the incoming beams is also used as a calibration of the driving amplitude and phase of the retro-reflecting mirror. From this we infer an uncertainty on the relative phase $\phi_{\text{r}}$ of at most $0.01\pi$, while the relative error on the amplitudes $K_n$ is $0.5\%$. An additional uncertainty for the amplitudes results from the imprecise knowledge of the site distance $d$ coming from the uncertainty of the lattice calibration. Due to a residual experimental mismatch of the phase modulations of the incoming and retro-reflected beams, the interference amplitude of the lattice is periodically reduced by at most $0.4\%$. To enter the driven regime, we linearly ramp up the modulation within $5.454\:\text{ms}$ and subsequently keep fixed driving amplitudes $K_1$ and $K_2$. For the gap measurements, we ramp the interactions while modulating the lattice from the initial value $U/h=5.41(7)\:\text{kHz}$ across the resonance to the final interaction $U/h=7.9(1)\:\text{kHz}$ within $20\:\text{ms}$. Alternatively, for the measurement of the tunnelling phase, we prepare an eigenstate of the resonantly driven double well by ramping the interactions to the resonance at $U/h=6.23(8)\:\text{kHz}$ within $100\:\text{ms}$. 

\subsection{Experimental measurement of the Peierls phase}
For the measurement of the tunnelling phase, we first prepare an eigenstate on the resonance $U=2\hbar\omega+\Delta_0$ as described above, followed by a projection onto an off-resonantly driven double well by switching the internal state of the atoms with an RF pulse. To realize the near-resonant condition, we work with a $\{\uparrow,\downarrow\}$-pair of atoms in the $m_F=-9/2$ and $m_F=-5/2$ states which are strongly interacting. Afterwards, we switch to a $\{\uparrow,\rightarrow\}$-pair of atoms in the $m_F=-9/2$ and $m_F=-7/2$ states, which have a much weaker on-site interaction energy. Importantly, we have to simultaneously match the two resonance conditions for the interactions $U_{-5/2,-9/2}=2\hbar\omega+\Delta_0$ and $U_{-7/2,-9/2}=\Delta_0$ at the same strength of the magnetic offset field. For our lattice configuration and choices of $\omega/(2\pi)=2.75\:\text{kHz}$ and $\Delta_0/h=660(20)\:\text{Hz}$, this condition is fulfilled for a magnetic field of $B_{\text{res}}=210.82(2)\:\text{G}$. Here, the scattering lengths are $a_{-5/2,-9/2}=273(1)\; a_0$ and $a_{-7/2,-9/2}=25(2)\; a_0$ and the corresponding on-site interactions are $U_{-5/2,-9/2}/h=6.23(8)\:\text{kHz}$ and $U_{-7/2,-9/2}/h=0.56(5)\:\text{kHz}$. To switch between the two different regimes, we transfer the atoms from the $m_F=-5/2$ to the $m_F=-7/2$ state with a fidelity of $95(4)\%$ by applying an RF pulse with a duration of $9.5\;\mu\text{s}$ and a frequency of $48.692\:\text{MHz}$. After the  interaction quench, the quantum state will start to rotate around the new off-resonant Hamiltonian on the Bloch sphere with a frequency of $2\sqrt{2}|t_{\text{eff}}^{(0)}|/h$. To measure the Ramsey fringes, we fix the evolution time to $\tau=h/(8\sqrt{2}|t_{\text{eff}}^{(0)}|)$ where the rotation angle is equal to $\pi/2$. Since $|t_{\text{eff}}^{(0)}|$ is changing as a function of our driving parameters $K_1$, $K_2$ and $\phi_{\text{r}}$ (see Supp.\;Fig.\;\ref{fig:S3}), we have to adjust the timing for each choice of parameters. We do this experimentally by projecting a pure singlet state with the RF pulse onto the off-resonant Hamiltonian, which results in coherent oscillations between the singlet and double occupancy states. From these oscillation, we extract the $\pi/2$ time for a certain set of driving parameters and interpolate between them. 

\subsection{Fit of the Ramsey fringes}
For the fringes, we perform 3 independent measurements of the final double occupancy fraction for 7 different values of the common phase $\phi_{\text{c}}$ between 0 and $\pi$ (see Fig.\;\ref{fig:2}d). To extract the tunnelling phase $\psi_{\text{r}}$, we fit the resulting double occupancy fringe with a function $\mathcal{D}(\phi_{\text{c}})=A \sin{(2\phi_{\text{c}}+\psi_{\text{r}})}+o$, where the period is fixed to $\pi$. To estimate the error, we use a resampling method which assumes that the measurement results for each value of $\phi_{\text{c}}$ follow a normal distribution according to the measured values of the mean and standard deviation of $\mathcal{D}$. Afterwards, we randomly sample a value for the double occupancy fraction for each common phase and refit the dataset. We repeat this procedure $1500$ times while additionally varying the initialization values for the fit parameters $A$ and $o$ by $\pm 10\%$. The mean +($-$) standard deviation of the distribution of phases fitted on the resampled data is used as an upper (lower) bound for the fitted value of $\psi_{\text{r}}$ of the measured data, which is expressed in asymmetric error bars in Figs.\;\ref{fig:3}c and \ref{fig:4}d,e. The same resampling method is also employed to estimate the uncertainty on the center position of the Lorentzian fits that are used to determine the gap closing (see Figs.\;\ref{fig:3}b, \ref{fig:4}b and Supp.\;Fig.\;\ref{fig:S7}b). 

\subsection{Phase offset of the Ramsey fringes}
In the measurements of the tunnelling phase, we observe an overall offset, i.e. the phase of the Ramsey fringes is not vanishing for $\psi_{\text{r}}=0$. This can be explained by the evolution of the state during the adiabatic preparation of the eigenstate of $H_{\text{eff}}^{(2)}$. In particular, the relative phase between the singlet and double occupancy states is not only given by $-2\phi_{\text{c}}+\psi_{\text{r}}$, but it has an additional dynamical phase contribution in the lab frame given by $-2\omega\tau$. Therefore, in addition to the slow adiabatic following to the equator of the Bloch sphere (see Fig.\;\ref{fig:2}c), the state vector rotates at a frequency of $\omega/(2\pi)$ around the $z$-axis. Even when fixing the total preparation time of the eigenstate to a multiple of the driving period, any residual detuning from the resonance will lead to a modified rotation frequency and therefore to a finite phase accumulation up to the point at which the RF pulse is applied. Since the preparation takes hundreds of driving cycles, this phase offset can be significant. Furthermore, finite frequency effects and dynamics that depend on the exact launching protocol of the drive lead to additional phase shifts. To calibrate the resulting phase offset in the experiment, we take 4 reference Ramsey fringes for a single frequency drive with $\omega/(2\pi)=5.5\:\text{kHz}$, both for positive and negative site offsets $\Delta_0$. For this single-frequency drive, the non-trivial contribution $\psi_{\text{r}}$ vanishes, which allows us to directly measure the phase offset. From these measurements, we obtain an offset of $-0.15(4)\pi$ (uncertainty denotes the standard error). 

\subsection{Detection}
The detection of the double occupancy and singlet fractions is similar to earlier work \cite{Jordens2008,Greif2013}. To characterise the state of the atoms, we first freeze all dynamics by quickly ramping up the tunnelling barrier in the double well within $100\;\mu\text{s}$ to a $V_{\overline{X},X,Y,Z}=[30,0,40,30]\;E_{\text{R}}$ cubic lattice. We can detect double occupancies both for a $\{\uparrow,\downarrow\}$-pair and a $\{\uparrow,\rightarrow\}$-pair of atoms. For this, we ramp down the magnetic field below the Feshbach resonance of the $m_F=-9/2$ and $m_F=-7/2$ atoms. We then selectively transfer one of the atoms forming the double occupancy from the $m_F=-5/2$ to the $m_F=-7/2$ state (or vice versa) with an RF sweep by making use of the interaction shift. We can count the number of atoms in each $m_F$-state by applying a Stern-Gerlach pulse during a time-of-flight expansion followed by absorption imaging. To detect singlets, we apply a magnetic field gradient after the lattice freeze, which leads to coherent oscillations between the singlet and triplet state. After properly adjusting the evolution time, we detect the singlet state by merging two adjacent sites by going to a $V_{\overline{X},X,Y,Z}=[0,30,40,30]\;E_{\text{R}}$ checkerboard lattice. In this process, the singlet state will be adiabatically tranformed to a double occupancy in the final lattice, which we can detect as outlined above. 

\subsection{Theoretical treatment of the driven double well}
We perform both analytical and numerical studies of a double well subject to a two-frequency drive (for an analytical derivation of the effective Hamiltonian and tunnelling matrix element see supplementary material, for details about the numerical simulation see \cite{Desbuquois2017}). To calculate the numerical quasi-energy spectrum and the Floquet-eigenstates of the time-dependent problem (see Supp. Figs.\;\ref{fig:S1}a-c and \ref{fig:S5}e-h), we use a Trotter decomposition to compute the evolution operator over one modulation cycle. The content of double occupancy and singlet states for a given Floquet-eigenstate $\left|f_i\right\rangle$ ($i=1,...,4$) is then given by $\mathcal{D}_i=\left|\left\langle\uparrow\downarrow,0\right.\left|f_i\right\rangle\right|^2+\left|\left\langle0,\uparrow\downarrow\right.\left|f_i\right\rangle\right|^2$ and $\mathcal{S}_i=\left|\left\langle\text{s}\right.\left|f_i\right\rangle\right|^2$, respectively. In addition, we perform a numerical simulation of the full gap measurement protocol described above (see Supp. Figs.\;\ref{fig:S5}a-c and \ref{fig:S6}). These calculations capture the full time-dependence of the system, i.e. the drive at frequencies $\omega/(2\pi)$ and $2\omega/(2\pi)$ as well as the ramps of the driving amplitudes $K_1$ and $K_2$ and the interaction $U$. In detail, we initiate the system in a singlet state at $U=5.41\:\text{kHz}$ and first increase the amplitudes $K_1$ and $K_2$ of the two-frequency drive within $5.454\:\text{ms}$ to their final values at a fixed relative phase $\phi_{\text{r}}$ and $\phi_{\text{c}}=0$. Then, $U$ is ramped to $U=7.9\:\text{kHz}$ within $20\:\text{ms}$. To determine the double occupancy in the final state $\left|\varphi_{\text{fin}}\right\rangle$, we compute both overlaps with the double occupancy states $\left|\left\langle\uparrow\downarrow,0\right.\left|\varphi_{\text{fin}}\right\rangle\right|^2$ and $\left|\left\langle0,\uparrow\downarrow\right.\left|\varphi_{\text{fin}}\right\rangle\right|^2$, respectively. The results for the same driving parameters as in Fig.\;\ref{fig:3} (Fig.\;\ref{fig:4}a) are shown in Supp.\;Fig.\;\ref{fig:S5}a-c (Supp.\;Fig.\;\ref{fig:S6}). 

\subsection{Data availability}
All data files are available from the corresponding author upon request. Source Data for Figs.\;2-4 and Supp.\;Figs.\;1d and 7 are provided with the online version of the paper.

\subsection{Code availability}
The source code for the fit of the Ramsey fringes is available from the corresponding author upon request.

\clearpage

\makeatletter

\setcounter{section}{0}
\setcounter{subsection}{0}
\setcounter{figure}{0}
\setcounter{equation}{0}
\setcounter{table}{0}
\setcounter{NAT@ctr}{0}

\renewcommand{\figurename}[1]{Supp.\;Fig.\;}
\renewcommand{\theequation}{S\@arabic\c@equation}
\renewcommand{\thetable}{S\@arabic\c@table}

\renewcommand{\theHfigure}{SH\thefigure}
\renewcommand{\theHequation}{SH\theequation}
\renewcommand{\theHtable}{SH\thetable}

\renewcommand{\bibnumfmt}[1]{[S#1]}
\renewcommand{\citenumfont}[1]{S#1}

\makeatother

\section{\large SUPPLEMENTARY INFORMATION}
\section{Part A: Effective Hamiltonian and dynamical gauge fields resulting from a resonant two-frequency drive}

\subsection{Derivation of the effective many-body Hamiltonian}
In the following, we derive in more detail the effective Hamiltonian for a Fermi-Hubbard model, which is driven at two frequencies that are resonant with the onsite interaction $U$ (see also Methods). This scheme allows for an independent control of both the amplitude and phase of the induced density-assisted hoppings with respect to the single-particle tunneling. For simplicity, we discuss the case of a one-dimensional lattice.

We start from the full time-dependent Hamiltonian for interacting fermions in a driven lattice, which can be written as
\begin{equation}
\hat{H}(\tau)=\hat{H}_0+\hat{V}(\tau).
\label{drivenFHM}
\end{equation}
Here, the static part is given by the usual Fermi-Hubbard model
\begin{equation}
\hat{H}_0=-t \sum_{j,\sigma} (\hat{c}^{\dagger}_{j+1,\sigma}\hat{c}_{j\sigma}+\text{h.c.})+U \sum_j \hat{n}_{j\uparrow}\hat{n}_{j\downarrow},
\label{staticFHM}
\end{equation}
where $\hat{c}^{\dagger}_{j\sigma}$ and $\hat{c}_{j\sigma}$ create and annihilate a fermion in spin state $\sigma\in\{\uparrow, \downarrow\}$ at site $j$, respectively, and $\hat{n}_{j\sigma}$ is the number operator. The drive is described by the operator
\begin{equation}
\hat{V}(\tau)=-f(\tau)\sum_{j,\sigma} j \hat{n}_{j\sigma},
\label{DriveOpManyBody}
\end{equation}
which corresponds to an inertial force that is sinusoidally modulated at two frequencies $\omega/(2\pi)$ and $2\omega/(2\pi)$
\begin{eqnarray}
f(\tau)&=&\hbar\omega K_1\cos(\omega\tau+\phi_{\text{c}})\notag\\
			 &+& 2\hbar\omega K_2\cos(2\omega\tau+2\phi_{\text{c}}+\phi_{\text{r}}).
\label{DriveTwoFreqManyBody}
\end{eqnarray}
Here, $\phi_{\text{c}}$ denotes the common phase, which shifts the waveform in time and $\phi_{\text{r}}$ is a relative phase. The latter explicitly breaks time-reversal (TR) symmetry of the waveform in the case that $\phi_{\text{r}}\neq 0,\pi$ (see Fig.\;\ref{fig:1}d). The parameters $K_1$ and $K_2$ are the dimensionless amplitudes of the two drives.

We now apply Floquet theory to derive an effective static Hamiltonian that describes the dynamics of the system on long timescales for interactions around the resonance $U\approx l\hbar\omega$ ($l\in\mathbb{Z}$)\;\cite{sBukov2015}. For this, we first go to an appropriate rotating frame in which we perform a high-frequency expansion. Since we have to take care of all resonances appearing in the system, we apply the transformation 
\begin{equation}
\hat{R}^{(l)}(\tau)=\exp\left[-i \left( l\omega\tau \sum_j \hat{n}_{j\uparrow}\hat{n}_{j\downarrow} - F(\tau)\sum_{j,\sigma} j \hat{n}_{j\sigma}\right)\right]
\label{TrafoRotTwoFreq}
\end{equation}
with
\begin{eqnarray}
F(\tau)&=& \frac{1}{\hbar}\int f(\tau)\text{d}\tau \notag\\
&=& K_1\sin(\omega\tau+\phi_{\text{c}})+K_2\sin(2\omega\tau+2\phi_{\text{c}}+\phi_{\text{r}}).\notag\\
\end{eqnarray}
The Hamiltonian transforms according to
\begin{eqnarray}
\hat{H}_{\text{rot}}^{(l)}&=& [\hat{R}^{(l)}]^{\dagger}\hat{H} \hat{R}^{(l)} - i\hbar [\hat{R}^{(l)}]^{\dagger}\frac{\partial}{\partial\tau}\hat{R}^{(l)} \label{rotFrame} \\
				&=& [\hat{R}^{(l)}]^{\dagger}\hat{H}_0 \hat{R}^{(l)},
\end{eqnarray}
where the explicit time-dependencies were omitted for clarity. With this we obtain the Hamiltonian in the rotating frame
\begin{eqnarray}
\hat{H}^{(l)}_{\text{rot}}(\tau)=&-&t \sum_{j,\sigma} \left[ e^{-iF(\tau)} e^{i l\omega\tau(\hat{n}_{j+1,\bar{\sigma}}-\hat{n}_{j\bar{\sigma}})}\hat{c}^{\dagger}_{j+1,\sigma}\hat{c}_{j\sigma} \right. \notag\\
&+& \left. \text{h.c.} \right] + (U-l\hbar\omega) \sum_j \hat{n}_{j\uparrow}\hat{n}_{j\downarrow}
\label{drivenFHMrot}
\end{eqnarray}
with $\bar{\uparrow}=\downarrow$ and vice versa. Here, it becomes explicit that the phase that is picked up during the tunnelling process depends on the number operator $\hat{n}_{j\bar{\sigma}}$. In addition, the effective interaction is given by the detuning from resonance $\delta^{(l)}=U-l\hbar\omega$.

Next, we separate single-particle and density-assisted hopping processes by introducing the operators
\begin{eqnarray}
\hat{a}_{\langle i,j \rangle,\sigma} &=& (1-\hat{n}_{i\sigma})(1-\hat{n}_{j\sigma})+\hat{n}_{i\sigma}\hat{n}_{j\sigma} \notag \\
\hat{b}_{\langle i,j \rangle,\sigma\text{R}} &=& (1-\hat{n}_{i\sigma})\hat{n}_{j\sigma} \label{ab} \\
\hat{b}_{\langle i,j \rangle,\sigma\text{L}} &=& \hat{n}_{i\sigma}(1-\hat{n}_{j\sigma}). \notag 
\end{eqnarray}
The operator $\hat{a}_{\langle j,j+1 \rangle,\sigma}$ describes processes where atoms tunnel between nearest-neighbor sites with equal occupations $n_{j+1,\bar{\sigma}}=n_{j\bar{\sigma}}$, while for $\hat{b}_{\langle j,j+1 \rangle,\sigma\text{R}}$ ($\hat{b}_{\langle j,j+1 \rangle,\sigma\text{L}}$) we have $n_{j+1,\bar{\sigma}}-n_{j\bar{\sigma}}=+ 1$ ($-1$) and the occupation is higher on the right (R) (left (L)) site, respectively. We can then write the Hamiltonian in Eq.\;(\ref{drivenFHMrot}) as
\begin{eqnarray}
\hat{H}^{(l)}_{\text{rot}}(\tau)=&-&t \sum_{j,\sigma} \left[ e^{-iF(\tau)} \left(\hat{a}_{\langle j,j+1 \rangle,\bar{\sigma}}+ e^{il\omega\tau}\hat{b}_{\langle j,j+1 \rangle,\bar{\sigma}\text{R}} \right. \right. \notag\\
&+& \left. \left. e^{-il\omega\tau}\hat{b}_{\langle j,j+1 \rangle,\bar{\sigma}\text{L}}\right) \hat{c}^{\dagger}_{j+1,\sigma}\hat{c}_{j\sigma}+\text{h.c.}\right] \notag\\
&+& (U-l\hbar\omega) \sum_j \hat{n}_{j\uparrow}\hat{n}_{j\downarrow}.
\label{drivenFHMrot2}
\end{eqnarray}
To lowest order, the effective Hamiltonian is given by the time average over one modulation period $\hat{H}^{(l)}_{\text{eff}}=\left\langle \hat{H}^{(l)}_{\text{rot}}(\tau)\right\rangle_{T}$ and we find
\begin{eqnarray}
\hat{H}^{(l)}_{\text{eff}}=&-&\sum_{j,\sigma} \left[\left(t_{\text{eff}}^{(0)}\hat{a}_{\langle j,j+1 \rangle,\bar{\sigma}}+ t_{\text{eff},\text{R}}^{(l)}\hat{b}_{\langle j,j+1 \rangle,\bar{\sigma}\text{R}} \right. \right. \notag\\
&+& \left. \left. \left[t_{\text{eff},\text{L}}^{(l)}\right]^* \hat{b}_{\langle j,j+1 \rangle,\bar{\sigma}\text{L}}\right) \hat{c}^{\dagger}_{j+1,\sigma}\hat{c}_{j\sigma}+\text{h.c.}\right] \notag\\
&+& (U-l\hbar\omega) \sum_j \hat{n}_{j\uparrow}\hat{n}_{j\downarrow}. 
\label{HeffMany}
\end{eqnarray}
Here, the single-particle hopping is described by $t_{\text{eff}}^{(0)}$, while the matrix elements $t_{\text{eff},\text{R/L}}^{(l)}$ correspond to density-assisted tunnellings, in which a double occupation is created by a hopping process to the right or left, respectively. The effective tunnellings are given by 
\begin{eqnarray}
t_{\text{eff},\text{R/L}}^{(l)} &=& te^{-il\phi_{\text{c}}}\sum_m \mathcal{J}_{-2m\pm l}(K_1)\mathcal{J}_m(K_2)e^{\mp im \phi_{\text{r}}} \notag \\
&=& |t_{\text{eff},\text{R/L}}^{(l)}|e^{i\psi_{\text{R/L}}^{(l)}} 
\label{teff}
\end{eqnarray}
for $l\in\mathbb{Z}$, where $\mathcal{J}_m$ is the $m$-th order Bessel function and we introduced the phase of the effective tunnelling $\psi_{\text{R/L}}^{(l)}=\text{Arg}(t_{\text{eff},\text{R/L}}^{(l)})$. This expression can be interpreted as the sum over all multi-photon processes in which $m$ photons are absorbed from the $2\omega$ drive and $2m\mp l$ photons are re-emitted into the $\omega$ drive, such that the total net energy added to the system is $l\hbar\omega$. For $K_2=0$ ($K_1=0$), the equation reduces to the usual density-assisted tunnelling amplitude of a single-frequency drive $\mathcal{J}_{\pm l}(K_1)e^{-il\phi_{\text{c}}}$ ($\mathcal{J}_{\pm l/2}(K_2)e^{-il(\phi_{\text{c}}+\phi_{\text{r}}/2)}$ with $l$ even).

The effective tunneling matrix elements are complex-valued and can be tuned independently in amplitude and phase. It can be shown that they obey the relation
\begin{equation}
t_{\text{eff},\text{L}}^{(l)}(\phi_{\text{c}},\phi_{\text{r}}) = t_{\text{eff},\text{R}}^{(l)}(\phi_{\text{c}}+\pi,\phi_{\text{r}}+\pi).
\label{tpm}
\end{equation}
In particular, this means that for resonances with $l$ odd, the sign of the tunnelling is different for tunnelling events which create a double occupancy on the left or right side. In addition, we find that
\begin{equation}
\left[t_{\text{eff},\text{R}}^{(l)}(\phi_{\text{c}},\phi_{\text{r}})\right]^* = t_{\text{eff},\text{R}}^{(l)}(-\phi_{\text{c}},-\phi_{\text{r}})
\label{teffConj}
\end{equation}
such that the tunnelling is real for the TR symmetric points $\phi_{\text{r}}=0$ and $\pi$ (up to a global phase that can be removed by a gauge transformation). Combining the expressions (\ref{tpm}) and (\ref{teffConj}), we can relate the tunneling matrix elements in the Hamiltonian (\ref{HeffMany}) via
\begin{equation}
\left[t_{\text{eff},\text{L}}^{(l)}(\phi_{\text{c}},\phi_{\text{r}})\right]^*=t_{\text{eff},\text{R}}^{(l)}(-\phi_{\text{c}}+\pi,-\phi_{\text{r}}+\pi).
\end{equation}
Specifically for the case that $l=0$ one can show the additional relation
\begin{equation}
t_{\text{eff},\text{R}}^{(0)}(\phi_{\text{c}}+\pi,\phi_{\text{r}}+\pi)=\left[t_{\text{eff},\text{R}}^{(0)}(\phi_{\text{c}},\phi_{\text{r}})\right]^*.
\label{teff0Pi}
\end{equation}
This means in particular that 
\begin{equation}
t_{\text{eff},\text{R}}^{(0)}(\phi_{\text{c}},\phi_{\text{r}}) = \left[t_{\text{eff},\text{L}}^{(0)}(\phi_{\text{c}},\phi_{\text{r}})\right]^*
\label{teff0LR}
\end{equation}
and the single-particle hopping in the Hamiltonian (\ref{HeffMany}) is fully described by the matrix element $t_{\text{eff}}^{(0)} \equiv t_{\text{eff},\text{R}}^{(0)}=\left[t_{\text{eff},\text{L}}^{(0)}\right]^*$. 

\subsection{Dynamical gauge fields and mapping on a link model}
In order to establish the connection of the effective Hamiltonian in Eq.\;(\ref{HeffMany}) to dynamical gauge fields, we rewrite it for the resonant case $U=l\hbar\omega$ in the general form
\begin{equation}
\hat{H}^{(l)}_{\text{eff}}=-\sum_{j,\sigma} \left(\hat{t}_{\langle j,j+1 \rangle,\bar{\sigma}}^{(l)} e^{i\hat{\mathcal{A}}_{\langle j,j+1 \rangle,\bar{\sigma}}^{(l)}} \hat{c}^{\dagger}_{j+1,\sigma}\hat{c}_{j\sigma}+\text{h.c.}\right).
\label{HeffManySpin}
\end{equation}
Here, it becomes apparent that atoms of the opposite spin state $\bar{\sigma}$ act as a link variable for the hopping of particles in state $\sigma$. In general, both the tunneling amplitude $\hat{t}_{\langle i,j \rangle,\bar{\sigma}}^{(l)}$ and phase $\hat{\mathcal{A}}_{\langle i,j \rangle,\bar{\sigma}}^{(l)}$ for $\sigma$ depend on the configuration of $\bar{\sigma}$ on the link $\langle i,j\rangle$ and are therefore operators. Using the operators in Eq.\;(\ref{ab}), they can be expressed as
\begin{eqnarray}
\hat{t}_{\langle i,j \rangle,\sigma}^{(l)} &=& |t_{\text{eff}}^{(0)}|\:\hat{a}_{\langle i,j \rangle,\sigma} + |t_{\text{eff},\text{R}}^{(l)}|\:\hat{b}_{\langle i,j \rangle,\sigma\text{R}} \notag \\
&+& |t_{\text{eff},\text{L}}^{(l)}|\:\hat{b}_{\langle i,j \rangle,\sigma\text{L}} \\
\hat{\mathcal{A}}_{\langle i,j \rangle,\sigma}^{(l)} &=& \psi^{(0)}\:\hat{a}_{\langle i,j \rangle,\sigma} + \psi_{\text{R}}^{(l)}\:\hat{b}_{\langle i,j \rangle,\sigma\text{R}} \notag \\
&-& \psi_{\text{L}}^{(l)}\:\hat{b}_{\langle i,j \rangle,\sigma\text{L}}.
\end{eqnarray}
The fact that $\hat{\mathcal{A}}_{\langle i,j \rangle,\sigma}^{(l)}$ is an operator gives rise to dynamical gauge fields in higher dimensions, since the magnetic flux that the atoms in state $\sigma$ experience when hopping around a plaquette
\begin{equation}
\hat{B}_{\Box,\bar{\sigma}}^{(l)}=\prod_{\langle \textbf{i},\textbf{j}\rangle \in \Box}  e^{i\hat{\mathcal{A}}_{\langle \textbf{i},\textbf{j}\rangle ,\bar{\sigma}}^{(l)}}
\label{DynamicalFlux}
\end{equation}
depends on the configuration of atoms in the opposite spin state $\bar{\sigma}$. This is different to the case of classical gauge fields, where the phase that the atoms pick up is independent from the other atoms. 

The link variables $\hat{t}_{\langle i,j \rangle,\sigma}^{(l)}$ and $\hat{\mathcal{A}}_{\langle i,j \rangle,\sigma}^{(l)}$ in the Hamiltonian (\ref{HeffManySpin}) can be represented in the eigenbasis of the pseudo spin-1 operator
\begin{equation}
\hat{\tau}_{\langle i,j \rangle,\sigma}^z = \hat{n}_{i\sigma}-\hat{n}_{j\sigma}, 
\label{PseudoSpinOps}
\end{equation}
which describes the occupation imbalance between sites $i$ and $j$ of atoms in state $\sigma$. The eigenstates of $\hat{\tau}_{\langle i,j \rangle,\sigma}^z$ fulfill the relation
\begin{equation}
\hat{\tau}_{\langle i,j \rangle,\sigma}^z \left|m_{\langle i,j \rangle,\sigma}\right\rangle = m_{\langle i,j \rangle,\sigma}\left|m_{\langle i,j \rangle,\sigma}\right\rangle
\end{equation}
with $m_{\langle i,j \rangle,\sigma}\in \{-1,0,1\}$. On each link $\langle i,j\rangle$ and for both spin states $\sigma$, the pseudo spin operator introduced in Eq.\;(\ref{PseudoSpinOps}) is then represented by the matrix
\begin{equation}
\hat{\tau}_{\langle i,j \rangle,\sigma}^z =\begin{pmatrix} 
							1 & 0 & 0 \\
							0 & 0 & 0 \\
							0 & 0 & -1 \end{pmatrix}.
\label{PseudoSpinOpsMatrix}
\end{equation}
On the other hand, the operators defined in Eq.\;(\ref{ab}) are the projections onto the eigenstates $\left|1\right\rangle$, $\left|0\right\rangle$ and $\left|-1\right\rangle$ given by
\begin{eqnarray}
&&\hat{b}_{\langle i,j \rangle,\sigma\text{L}} =\begin{pmatrix} 
							1 & 0 & 0 \\
							0 & 0 & 0 \\
							0 & 0 & 0 \end{pmatrix},\  
\hat{a}_{\langle i,j \rangle,\sigma} =\begin{pmatrix} 
							0 & 0 & 0 \\
							0 & 1 & 0 \\
							0 & 0 & 0 \end{pmatrix},\notag \\
&&\hat{b}_{\langle i,j \rangle,\sigma\text{R}} =\begin{pmatrix} 
							0 & 0 & 0 \\
							0 & 0 & 0 \\
							0 & 0 & 1 \end{pmatrix}.
\end{eqnarray}
Therefore, the operators correponding to the tunneling amplitude and phase in the Hamiltonian (\ref{HeffManySpin}) are represented by the matrices
\begin{equation}
\hat{t}_{\langle i,j \rangle,\sigma}^{(l)} = \begin{pmatrix} 
							|t_{\text{eff},\text{L}}^{(l)}| & 0 & 0 \\
							0 & |t_{\text{eff}}^{(0)}| & 0 \\
							0 & 0 & |t_{\text{eff},\text{R}}^{(l)}| \end{pmatrix} 
\label{TunnelOp}
\end{equation}
and
\begin{equation}
\hat{\mathcal{A}}_{\langle i,j \rangle,\sigma}^{(l)} = \begin{pmatrix} 
							-\psi_{\text{L}}^{(l)} & 0 & 0 \\
							0 & \psi^{(0)} & 0 \\
							0 & 0 & \psi_{\text{R}}^{(l)} \end{pmatrix},
\label{GaugeOp}
\end{equation}
respectively. 

Next, let us discuss possible gauge transformations that leave the effective Hamiltonian unchanged. First, we can perform a transformation corresponding to the unitary operator
\begin{equation}
\hat{R}_{1,\sigma}(\varphi_{\sigma})=\exp\left(-i \varphi_{\sigma} \sum_{j} j \hat{n}_{j\sigma}\right)
\label{GaugeTrafoR1}
\end{equation}
which acts on the spin state $\sigma$. This corresponds to the second part of the transformation to the rotating frame in Eq.\;(\ref{TrafoRotTwoFreq}) and changes the gauge field operators according to
\begin{eqnarray}
\hat{\mathcal{A}}_{\langle i,j \rangle,\sigma}^{(l)} &&\rightarrow \hat{\mathcal{A}}_{\langle i,j \rangle,\sigma}^{(l)}\\
\hat{\mathcal{A}}_{\langle i,j \rangle,\bar{\sigma}}^{(l)} &&\rightarrow \hat{\mathcal{A}}_{\langle i,j \rangle,\bar{\sigma}}^{(l)} + \varphi_{\sigma} \mathds{1}_{\langle i,j \rangle,\bar{\sigma}}. 
\end{eqnarray}
As intuitively expected, the operators $\hat{\mathcal{A}}_{\langle i,j \rangle,\sigma}^{(l)}$ are only defined up to a global phase, which appears equivalently for all tunneling processes. 

The second transformation that we can perform is described by the operator
\begin{equation}
\hat{R}_2(\varphi_z)=\exp\left(-i \varphi_z \sum_j \hat{n}_{j\uparrow}\hat{n}_{j\downarrow}\right),
\label{GaugeTrafoR2}
\end{equation}
which, importantly, acts simultaneously on both spin states. As we know from the transformation to the rotating frame in Eq.\;(\ref{TrafoRotTwoFreq}), this changes the gauge fields according to
\begin{eqnarray}
\hat{\mathcal{A}}_{\langle i,j \rangle,\uparrow}^{(l)} &&\rightarrow \hat{\mathcal{A}}_{\langle i,j \rangle,\uparrow}^{(l)} - \varphi_z \hat{\tau}_{\langle i,j \rangle,\uparrow}^z \\
\hat{\mathcal{A}}_{\langle i,j \rangle,\downarrow}^{(l)} &&\rightarrow \hat{\mathcal{A}}_{\langle i,j \rangle,\downarrow}^{(l)}  - \varphi_z \hat{\tau}_{\langle i,j \rangle,\downarrow}^z. 
\end{eqnarray}
In other words, it is possible to gauge away phases which atoms in both spin states experience and that appear equally for the two density-assisted tunneling processes to the left and right. By comparing the expression to Eq.\;(\ref{teff}), we can identify $\varphi_z=-l\phi_{\text{c}}$. Therefore, such phases correspond to different common phases of the driving waveform in Eq.\;(\ref{DriveTwoFreqManyBody}), and the operator (\ref{GaugeTrafoR2}) represents the Floquet gauge transformation that shifts the origin of time. 

\subsection{Two-frequency driving schemes to obtain $\mathbb{Z}_N$ gauge fields}
In lattice gauge theories, the matter field is represented by particles that can hop between the vertices of the lattice, while the gauge particles are located on the bonds and act as link variables. We now want to discuss how the absolute values and phases of the effective tunnelings that we obtain with the two-frequency driving scheme (see Eqs.\;(\ref{TunnelOp}) and (\ref{GaugeOp})) have to be engineered in order that the effective Hamiltonian is reminiscent of a lattice gauge theory. For this, we need two main ingredients: 
\begin{enumerate}
	\item The symmetry between the two species ($\uparrow$ and $\downarrow$) has to be broken, such that atoms of one species (say $\uparrow$) can be identified as the matter particles, which experience a flux from the other species ($\downarrow$) representing the gauge field. In other words, we want to engineer gauge fields of the form
\begin{eqnarray}
\hat{\mathcal{A}}_{\langle i,j \rangle,\downarrow}^{(l)}&=&\text{diag}\left(-\psi_{\text{L}\uparrow}^{(l)},\psi_{\uparrow}^{(0)},\psi_{\text{R}\uparrow}^{(l)}\right) \label{GaugeOpAsymmetric} \\
\hat{\mathcal{A}}_{\langle i,j \rangle,\uparrow}^{(l)}&=&\mathbf{0}. 
\end{eqnarray}

\item In order to simplify the Hamiltonian and to be dominated by the effects of the tunneling phases, the hopping amplitudes should be independent of the site occupations, i.e. $|t_{\text{eff},\text{L}}^{(l)}|=|t_{\text{eff}}^{(0)}|=|t_{\text{eff},\text{R}}^{(l)}|\equiv t_{\text{eff}}$. In this case, the tunneling operators $\hat{t}_{\langle i,j \rangle,\sigma}^{(l)}$ in Eq.\;(\ref{TunnelOp}) are given by
\begin{eqnarray}
\hat{t}_{\langle i,j \rangle,\downarrow}^{(l)} &=& t_{\text{eff},\uparrow}\mathds{1}_{\langle i,j \rangle,\downarrow} \label{TunOpDownCnumber} \\
\hat{t}_{\langle i,j \rangle,\uparrow}^{(l)} &=& t_{\text{eff},\downarrow}\mathds{1}_{\langle i,j \rangle,\uparrow} \label{TunOpUpCnumber}
\end{eqnarray}
and effectively become $\mathbb{C}$-numbers. 
\end{enumerate}
If these two conditions are fulfilled, the effective Hamiltonian in Eq.\;(\ref{HeffManySpin}) is given by 
\begin{eqnarray}
\hat{H}^{(l)}_{\text{eff}}= &-&t_{\text{eff},\uparrow} \sum_{j} \left(e^{i\hat{\mathcal{A}}_{\langle j,j+1 \rangle,\downarrow}^{(l)}} \hat{c}^{\dagger}_{j+1,\uparrow}\hat{c}_{j\uparrow}+\text{h.c.}\right)\notag \\
&-&t_{\text{eff},\downarrow} \sum_{j} (\hat{c}^{\dagger}_{j+1,\downarrow}\hat{c}_{j\downarrow} + \text{h.c.}), 
\label{HeffManyLGT}
\end{eqnarray}
with the gauge field operator $\hat{\mathcal{A}}_{\langle i,j \rangle,\downarrow}^{(l)}$ in Eq.\;(\ref{GaugeOpAsymmetric}). In the following, we will outline how we can fulfill the two conditions above with a two-frequency driving scheme, which allows to engineer a tunable gauge field operator $\hat{\mathcal{A}}_{\langle i,j \rangle,\downarrow}^{(l)}$. 

In order to break the symmetry between the two spin states $\uparrow$ and $\downarrow$, we can slightly modify the driving scheme in Eq.\;(\ref{DriveOpManyBody}) and use a state selective two-frequency modulation of the form
\begin{eqnarray}
\hat{V}(\tau)&=&-\hbar\omega\sum_{j} j\left[ K_{\downarrow}\cos(\omega\tau+\phi_{\text{c}}) \hat{n}_{j\downarrow} \right. \notag \\
&&+ \left. n K_{\uparrow}\cos(n\omega\tau+n\phi_{\text{c}}+\phi_{\text{r}}) \hat{n}_{j\uparrow} \right].
\label{DriveOpManyBodySpinDep} 
\end{eqnarray}
Here, the two spin states are driven either at a frequency $\omega/(2\pi)$ with amplitude $K_{\downarrow}$ (for the gauge particle $\downarrow$) or at its multiple $n\omega/(2\pi)$ with amplitude $K_{\uparrow}$ (for the matter particle $\uparrow$), where $n\in\mathbb{N}$. Spin-dependent drives have been realized experimentally in cold atom setups by modulating a magnetic field gradient \cite{sJotzu2015}. Alternatively, they can be implemented by using two atomic species with different masses, which lead to distinct inertial forces when modulating the lattice position. As before, the modulation frequencies are chosen in resonance with the interactions $U=l\hbar\omega$ ($l\in\mathbb{Z}$). 

This driving scheme breaks the symmetry between the two internal states and allows to fully tune the dynamical gauge field experienced by the $\uparrow$ atoms. Furthermore, the amplitudes of the single-particle and density-assisted tunnelings can be matched to obtain the operators (\ref{TunOpDownCnumber}) and (\ref{TunOpUpCnumber}) by choosing appropriate driving strengths $K_{\sigma}$. We will illustrate this for two concrete examples for which we obtain a $\mathbb{Z}_2$ or $\mathbb{Z}_3$ gauge field. 

\subsubsection{$\mathbb{Z}_2$ gauge field}
If we choose the same frequency conditions as in our experiments (see Eq.\;(\ref{DriveOpManyBody})) with $n=2$ and $l=2$, the effective tunnelings for the driving scheme (\ref{DriveOpManyBodySpinDep}) are given by (see Eq.\;(\ref{teff}))
\begin{equation}
\begin{aligned}
&t_{\uparrow}^{(0)} = \mathcal{J}_0(K_{\uparrow})\ \ \text{and} &t_{\uparrow,\text{R/L}}^{(2)} = &\pm \mathcal{J}_1(K_{\uparrow})e^{-i(2\phi_{\text{c}}+\phi_{\text{r}})} \\
&t_{\downarrow}^{(0)} = \mathcal{J}_0(K_{\downarrow})\ \ \text{and} &t_{\downarrow,\text{R/L}}^{(2)} = &\mathcal{J}_2(K_{\downarrow})e^{-2i\phi_{\text{c}}}. 
\end{aligned}
\end{equation}
This means that the tunneling operators $\hat{t}_{\langle i,j \rangle,\sigma} \equiv \hat{t}_{\langle i,j \rangle,\sigma}^{(2)}$ in Eq.\;(\ref{TunnelOp}) have the following representations
\begin{eqnarray}
\hat{t}_{\langle i,j \rangle,\downarrow} &=& \begin{pmatrix} 
							\mathcal{J}_1(K_{\uparrow}) & 0 & 0 \\
							0 & \mathcal{J}_0(K_{\uparrow}) & 0 \\
							0 & 0 & \mathcal{J}_1(K_{\uparrow}) \end{pmatrix} \notag \\
\hat{t}_{\langle i,j \rangle,\uparrow} &=& \begin{pmatrix} 
							\mathcal{J}_2(K_{\downarrow}) & 0 & 0 \\
							0 & \mathcal{J}_0(K_{\downarrow}) & 0 \\
							0 & 0 & \mathcal{J}_2(K_{\downarrow}) \end{pmatrix}. 
\label{TunOpZ2}
\end{eqnarray}
The single-particle and density-assisted tunneling amplitudes can be matched by choosing 
\begin{eqnarray}
t_{\text{eff},\uparrow} &=& t\mathcal{J}_0(K_{\uparrow}) = t\mathcal{J}_1(K_{\uparrow}) \\ 
t_{\text{eff},\downarrow} &=& t\mathcal{J}_0(K_{\downarrow}) = t\mathcal{J}_2(K_{\downarrow}). 
\end{eqnarray}
The lowest driving amplitudes for which these conditions are fulfilled are $K_{\uparrow}\approx 1.43$ and $K_{\downarrow}\approx 1.84$, for which $t_{\text{eff},\uparrow}/t\approx 0.55$ and $t_{\text{eff},\downarrow}/t\approx 0.32$. In this case, the operators in Eq.\;(\ref{TunOpZ2}) reduce to the ones in Eqs.\;(\ref{TunOpDownCnumber}) and (\ref{TunOpUpCnumber}), respectively. 
Furthermore, the gauge field operators $\hat{\mathcal{A}}_{\langle i,j \rangle,\sigma}\equiv \hat{\mathcal{A}}_{\langle i,j \rangle,\sigma}^{(2)}$ in Eq.\;(\ref{GaugeOp}) are given by
\begin{equation}
\hat{\mathcal{A}}_{\langle i,j \rangle,\downarrow} = \begin{pmatrix} 
							\phi_{\text{r}}+\pi & 0 & 0 \\
							0 & 0 & 0 \\
							0 & 0 & -\phi_{\text{r}} \end{pmatrix},\ 
\hat{\mathcal{A}}_{\langle i,j \rangle,\uparrow} = \mathbf{0}
\label{GaugeOpZ2}
\end{equation}
where we chose a gauge in which $\varphi_{\downarrow}=\varphi_{\uparrow}=0$ and $\phi_{\text{c}}=0$ (see Eqs.\;(\ref{GaugeTrafoR1}) and (\ref{GaugeTrafoR2})). The phase $\phi_{\text{r}}$ is fully tunable from $0$ to $2\pi$ and non-trivial, i.e. it cannot be eliminated by a gauge transformation. In particular, although applying the transformation (\ref{GaugeTrafoR2}) with $\varphi_z=-\phi_{\text{r}}$ eliminates the phase from the operator $\hat{\mathcal{A}}_{\langle i,j \rangle,\downarrow}$, it reallocates it to the gauge field $\hat{\mathcal{A}}_{\langle i,j \rangle,\uparrow}$. In other words, the phase $\phi_{\text{r}}$ can be redistributed from the matter to the gauge particles. 

In such a model, it is for example straightforward to realize a $\mathbb{Z}_2$ lattice gauge theory. To this end, we restrict the number of gauge particles to one on each link $n_{j\downarrow}+n_{j+1,\downarrow}=1$, such that only density-assisted tunneling processes occur. In this case, the effective Hilbert space for the link variables $\downarrow$ is a two level system corresponding to the gauge particle sitting on site $j$ or $j+1$. If we choose $\phi_{\text{r}}=-\pi$, we obtain the dynamical $\mathbb{Z}_2$ gauge field
\begin{equation}
\hat{\mathcal{A}}_{\langle i,j \rangle,\downarrow}=\begin{pmatrix} 
							0 & 0 \\
							0 & \pi \end{pmatrix}
\end{equation}
such that 
\begin{equation}
e^{i\hat{\mathcal{A}}_{\langle i,j \rangle,\downarrow}} = \hat{\tau}_{\langle j,j+1 \rangle,\downarrow}^z \equiv \begin{pmatrix} 
							1 & 0 \\
							0 & -1 \end{pmatrix}. 
\end{equation}
On the other hand, the state of the link variable can be changed by the tunneling of gauge particles and we identify
\begin{equation}
\hat{c}^{\dagger}_{j+1,\downarrow}\hat{c}_{j\downarrow} + \text{h.c.} = \hat{\tau}_{\langle j,j+1 \rangle,\downarrow}^x \equiv \begin{pmatrix} 
							0 & 1 \\
							1 & 0 \end{pmatrix}. 
\end{equation}
Therefore, for this choice of driving parameters the effective Hamiltonian in Eq.\;(\ref{HeffManyLGT}) is given by 
\begin{eqnarray}
\hat{H}_{\mathbb{Z}_2}= &-&t_{\text{eff},\uparrow} \sum_{j}  \hat{\tau}_{\langle j,j+1 \rangle,\downarrow}^z (\hat{c}^{\dagger}_{j+1,\uparrow}\hat{c}_{j\uparrow}+\text{h.c.}) \notag\\
&-&t_{\text{eff},\downarrow} \sum_{j} \hat{\tau}_{\langle j,j+1 \rangle,\downarrow}^x.  
\end{eqnarray}
This model maps onto a lattice gauge theory featuring a $\mathbb{Z}_2$ symmetry \cite{sBarbiero2018}. In particular, the Pauli matrix $\hat{\tau}_{\langle i,j \rangle,\downarrow}^x$ can be identified as the $\mathbb{Z}_2$ electric field and $\hat{Q}_{j\uparrow}=\exp(i\pi\hat{n}_{j\uparrow})$ as the local $\mathbb{Z}_2$ charges. In higher dimensions, the matter particles $\uparrow$ acquire dynamical magnetic fluxes
\begin{equation}
\hat{B}_{\Box,\downarrow}=\prod_{\langle \textbf{i},\textbf{j}\rangle \in \Box}  
\hat{\tau}_{\langle \textbf{i},\textbf{j}\rangle,\downarrow}^z
\end{equation}
(see Eq.\;(\ref{DynamicalFlux})). 

\subsubsection{$\mathbb{Z}_3$ gauge field}
As another example, we consider the case $n=1$ and $l=2$ for the driving protocol in Eq.\;(\ref{DriveOpManyBodySpinDep}), i.e. the two internal states are modulated at the same frequency but at different phases. In this case, the effective tunneling matrix elements are given by
\begin{equation}
\begin{aligned}
&t_{\uparrow}^{(0)} = \mathcal{J}_0(K_{\uparrow})\ \ \text{and} &t_{\uparrow,\text{R/L}}^{(2)} = &\mathcal{J}_2(K_{\uparrow})e^{-2i(\phi_{\text{c}}+\phi_{\text{r}})} \\
&t_{\downarrow}^{(0)} = \mathcal{J}_0(K_{\downarrow})\ \ \text{and} &t_{\downarrow,\text{R/L}}^{(2)} = &\mathcal{J}_2(K_{\downarrow})e^{-2i\phi_{\text{c}}}.
\end{aligned}
\end{equation}
Hence, if we do not restrict the number of gauge particles on each link, the tunneling operators in Eq.\;(\ref{TunnelOp}) have the representations
\begin{equation}
\hat{t}_{\langle i,j \rangle,\sigma} = \begin{pmatrix} 
							\mathcal{J}_2(K_{\bar{\sigma}}) & 0 & 0 \\
							0 & \mathcal{J}_0(K_{\bar{\sigma}}) & 0 \\
							0 & 0 & \mathcal{J}_2(K_{\bar{\sigma}}) \end{pmatrix}.
\end{equation}
Again, we can choose
\begin{equation}
t_{\text{eff},\sigma} = t\mathcal{J}_0(K_{\sigma}) = t\mathcal{J}_2(K_{\sigma}) 
\end{equation}
by setting $K_{\uparrow}=K_{\downarrow}\approx 1.84$, such that the tunneling operators reduce to the ones in Eqs.\;(\ref{TunOpDownCnumber}) and (\ref{TunOpUpCnumber}) with $t_{\text{eff},\uparrow}/t=t_{\text{eff},\downarrow}/t \approx 0.32$. 

Moreover, the gauge field operators $\hat{\mathcal{A}}_{\langle i,j \rangle,\sigma}\equiv \hat{\mathcal{A}}_{\langle i,j \rangle,\sigma}^{(2)}$ in Eq.\;(\ref{GaugeOp}) are given by
\begin{equation}
\hat{\mathcal{A}}_{\langle i,j \rangle,\downarrow} = \begin{pmatrix} 
							2\phi_{\text{r}} & 0 & 0 \\
							0 & 0 & 0 \\
							0 & 0 & -2\phi_{\text{r}} \end{pmatrix},\ 
\hat{\mathcal{A}}_{\langle i,j \rangle,\uparrow} = \mathbf{0}
\end{equation}
in the gauge $\varphi_{\downarrow}=\varphi_{\uparrow}=0$ and $\phi_{\text{c}}=0$. As in the operator (\ref{GaugeOpZ2}), the phase $\phi_{\text{r}}$ cannot be eliminated by a gauge transformation and is fully tunable. In particular, if we choose $\phi_{\text{r}}=-\pi/3$, we obtain a $\mathbb{Z}_3$ gauge field
\begin{equation}
\hat{\mathcal{A}}_{\langle i,j \rangle,\downarrow}=\begin{pmatrix} 
							0 & 0 & 0 \\
							0 & \frac{2\pi}{3} & 0 \\
							0 & 0 & \frac{4\pi}{3} \end{pmatrix} 
\end{equation}
in the gauge with $\varphi_{\uparrow}=2\pi/3$. Tuning the phase $\phi_{\text{r}}$ from $0$ to $-\pi/3$ allows to continuously interpolate between the trivial case $\hat{\mathcal{A}}_{\langle i,j \rangle,\downarrow}= \mathbf{0}$ to a dynamical $\mathbb{Z}_3$ gauge field. 

Furthermore, the tunneling of the gauge particles is represented by the matrix
\begin{equation}
\hat{c}^{\dagger}_{j+1,\downarrow}\hat{c}_{j\downarrow} + \text{h.c.}
\equiv \begin{pmatrix} 
							0 & 0 & 1 \\
							0 & 0 & 0 \\
							1 & 0 & 0 \end{pmatrix},
\end{equation}
i.e. it connects the spin states with eigenvalues $m_{\langle i,j \rangle,\downarrow}=\pm 1$ on each link. Instead of using a fermion in state $\downarrow$, it is also feasible to use bosonic atoms to represent the gauge particles, which are described by the bosonic creation and annihilation operators $\hat{b}^{\dagger}_j$ and $\hat{b}_j$, respectively. In this case, the form of the Hamiltonian in the rotating frame (Eq.\;(\ref{drivenFHMrot})) is still valid when replacing the operators for the fermion in state $\downarrow$ by the respective boson operators. However, it is now possible to work with more than one boson on each link, which allows to access higher dimensional Hilbert spaces. For example, when using two bosons on the links, we can again define a spin-1 operator as the occupation imbalance (compare to Eq.\;(\ref{PseudoSpinOps}))
\begin{equation}
\hat{\tau}_{\langle i,j \rangle}^z = \frac{\hat{n}^{\text{b}}_i-\hat{n}^{\text{b}}_j}{2},
\end{equation}
where $\hat{n}^{\text{b}}_j=\hat{b}^{\dagger}_j \hat{b}_j$. In this case, all three states of the link variable are connected via the tunneling of bosons
\begin{equation}
\hat{b}^{\dagger}_{j+1}\hat{b}_{j}  + \text{h.c.} = \hat{\tau}_{\langle j,j+1 \rangle}^x \equiv \begin{pmatrix} 
							0 & 1 & 0 \\
							1 & 0 & 1 \\
							0 & 1 & 0 \end{pmatrix}. 
\end{equation}
This approach can be readily extended to $N>3$ dimensional Hilbert spaces on the links by using $N-1$ bosons to represent the gauge particles.

\section{Part B: Two-site Hamiltonian and properties of the tunneling matrix element}
\subsection{Projection of the effective Hamiltonian on a double well}
After deriving the effective many-body Hamiltonian for the two-frequency driving scheme, we now project it onto a double well. In this system we can devise schemes that allow us to directly measure both the amplitude and Peierls phase of the effective tunneling matrix elements given in Eq.\;(\ref{teff}). 

We investigate a double well system with two distinguishable, interacting Fermions (labeled $\uparrow$ and $\downarrow$) which is driven at two frequencies (for a detailed analysis for the case of a single frequency refer to \cite{sDesbuquois2017}, Appendix A). 
As in the many-body case, the time-dependent Hamiltonian is given by
\begin{equation}
\hat{H}(\tau)=\hat{H}_0+\hat{V}(\tau). 
\end{equation}
Here, the static part contains the tunnel coupling $t$, the on-site interaction $U$ and a static energy bias between the two sites $\Delta_0$
\begin{equation}
\hat{H}_0=-\sqrt{2}\:t (\hat{H}_t+ \text{h.c.}) + U \hat{H}_U + \Delta_0 \hat{H}_{\Delta}. 
\label{Hstatic}
\end{equation}
As basis states, we choose the double occupancy states $\left|\uparrow\downarrow,0\right\rangle$ and $\left|0,\uparrow\downarrow\right\rangle$, where both particles are located on the left or right site, respectively, and the singlet and triplet states 
\begin{eqnarray}
\left|\text{s}\right\rangle=\left(\left|\uparrow,\downarrow\right\rangle-\left|\downarrow,\uparrow\right\rangle\right)/\sqrt{2} \\
\left|\text{t}\right\rangle=\left(\left|\uparrow,\downarrow\right\rangle+\left|\downarrow,\uparrow\right\rangle\right)/\sqrt{2}.
\end{eqnarray}
In this basis, the operators in Eq.\;(\ref{Hstatic}) can be represented as
\begin{eqnarray}
\hat{H}_t &=& \left|\uparrow\downarrow,0\right\rangle\left\langle \text{s}\right| + \left|0,\uparrow\downarrow\right\rangle\left\langle \text{s}\right| \notag\\
\hat{H}_U &=& \left|\uparrow\downarrow,0\right\rangle\left\langle \uparrow\downarrow,0\right| + \left|0,\uparrow\downarrow\right\rangle\left\langle 0,\uparrow\downarrow\right| \\
\hat{H}_{\Delta} &=& \left|\uparrow\downarrow,0\right\rangle\left\langle \uparrow\downarrow,0\right| - \left|0,\uparrow\downarrow\right\rangle\left\langle 0,\uparrow\downarrow\right|. \notag
\end{eqnarray}
Note that the triplet state does not couple to any other state and always remains an eigenstate at zero energy. The time-dependent part consists of a sinusoidal modulation of the site offset at two frequencies $\omega/(2\pi)$ and $2\omega/(2\pi)$
\begin{eqnarray}
\hat{V}(\tau)&=& \Delta(\tau) \hat{H}_{\Delta} \\
			 &=&\left[\hbar\omega K_1\cos(\omega\tau+\phi_{\text{c}})\right.\notag\\
			 &&+\left. 2\hbar\omega K_2\cos(2\omega\tau+2\phi_{\text{c}}+\phi_{\text{r}}) \right] \hat{H}_{\Delta}
\label{Drive}
\end{eqnarray}
(compare to Eq.\;(\ref{DriveTwoFreqManyBody})). 

As in the many-body case, we now apply Floquet theory to derive an effective static Hamiltonian that is valid around the resonance $U\approx l\hbar\omega$ ($l\in\mathbb{Z}$). We go to a rotating frame via the transformation in Eq.\;(\ref{TrafoRotTwoFreq}), which can be written as
\begin{eqnarray}
\hat{R}^{(l)}(\tau)&=& \exp\left\{-i l\omega\tau \hat{H}_U-\frac{i}{\hbar} \int{\hat{V}(\tau)d\tau}\right\} \\
			 &=& \exp\left\{-i l\omega\tau \hat{H}_U-i[K_1\sin(\omega\tau+\phi_{\text{c}})\right.\notag\\
			 && +\left.K_2\sin(2\omega\tau+2\phi_{\text{c}}+\phi_{\text{r}})]\hat{H}_{\Delta}\right\}.
\label{RotFrameTransf}
\end{eqnarray}
The Hamiltonian transforms according to Eq.\;(\ref{rotFrame}) and is given by
\begin{eqnarray}
\hat{H}_{\text{rot}}^{(l)}(\tau) = &-&\sqrt{2}\left[t_{\text{L}}^{(l)}(\tau) \hat{H}_{\text{L}}+t_{\text{R}}^{(l)}(\tau) \hat{H}_{\text{R}}+\text{h.c.}\right]\notag\\
&+& (U-l\hbar\omega) \hat{H}_U + \Delta_0 \hat{H}_{\Delta},
\label{Hrot}
\end{eqnarray}
where the operators $\hat{H}_{\text{L}}=\left|\uparrow\downarrow,0\right\rangle\left\langle \text{s}\right|$ and $\hat{H}_{\text{R}}=\left|0,\uparrow\downarrow\right\rangle\left\langle \text{s}\right|$ describe the coupling of the singlet state to a double occupancy on the left or right site, respectively, and the corresponding density-assisted tunnelling matrix elements are given by
\begin{eqnarray}
t_{\text{R/L}}^{(l)}(\tau)&=& t\exp\left\{i[l \omega\tau \mp K_1\sin(\omega\tau+\phi_{\text{c}})\right.\notag\\
			 && \mp\left.K_2\sin(2\omega\tau+2\phi_{\text{c}}+\phi_{\text{r}})]\right\}.
\end{eqnarray}
To lowest order, the effective Hamiltonian is given by the time average over one period $T=2\pi/\omega$ and can be described by an effective tunnelling matrix element $t_{\text{eff},\text{R/L}}^{(l)}=\left\langle t_{\text{R/L}}^{(l)}(\tau)\right\rangle_{T}$ given in Eq.\;(\ref{teff}). 

\begin{figure*}
	\includegraphics[width=178mm]{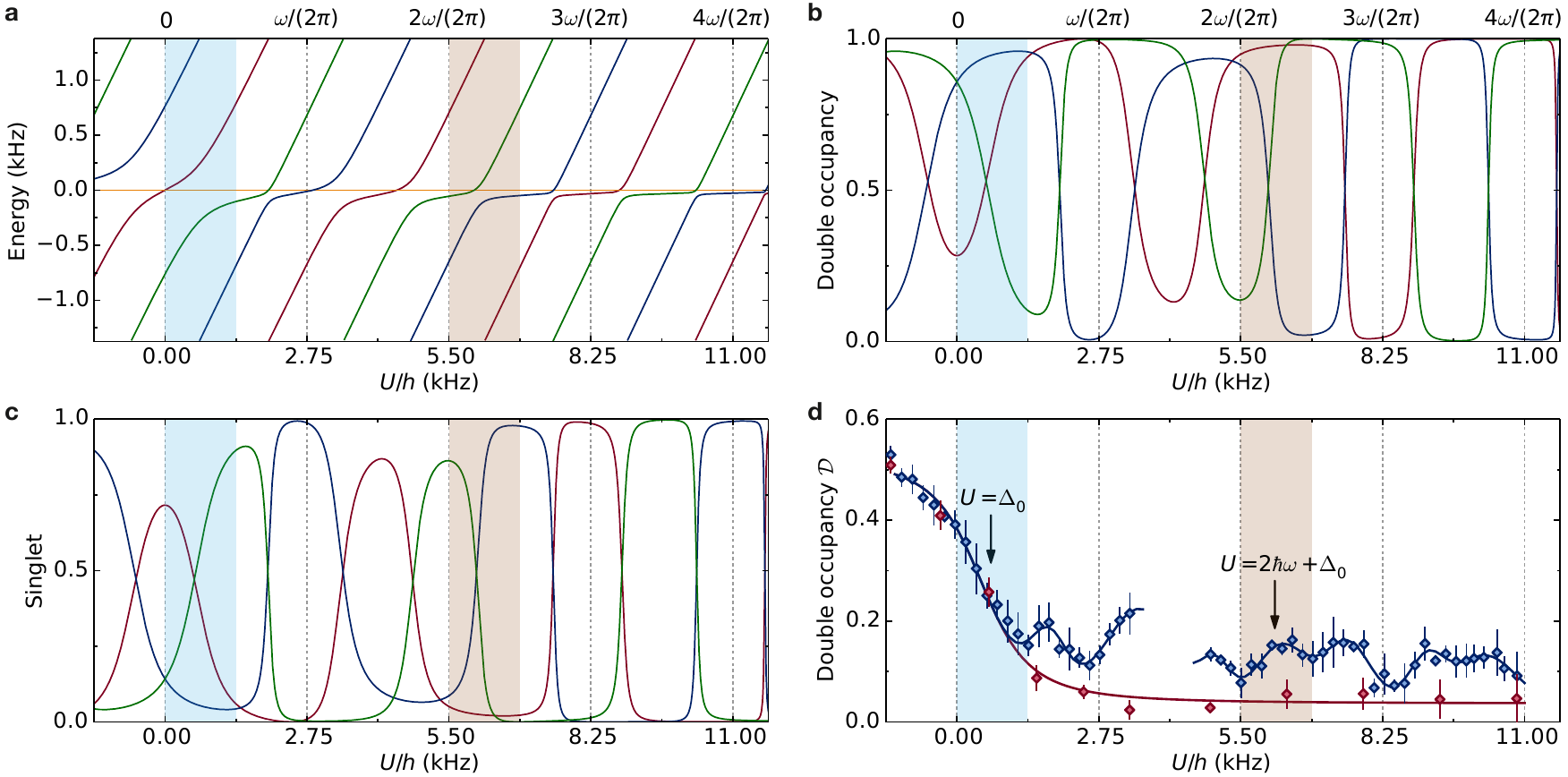}
	\caption{
		\textbf{Resonances in the double well system subject to a two-frequency drive.}
		\textbf{a-c,}
		Theoretical spectrum and eigenstates in a double well with the experimental tight-binding parameters $t/h=260\:\text{Hz}$ and $\Delta_0/h=660\:\text{Hz}$ as a function of $U$ for a two-frequency drive with $\omega/(2\pi)=2.75\:\text{kHz}$, $2\omega/(2\pi)=5.5\:\text{kHz}$, $K_1=K_2=0.79$ and $\phi_{\text{r}}=\pi$. The quasi-energy spectrum in \textbf{a} shows multiple avoided crossings at the resonances $U=l\hbar\omega\pm\Delta_0$ with gaps given by $2\sqrt{2}|t_{\text{eff,R/L}}^{(l)}|$. At each resonance, a double occupancy state ($\left|0,\uparrow\downarrow\right\rangle$ for $U=l\hbar\omega+\Delta_0$ and $\left|\uparrow\downarrow,0\right\rangle$ for $U=l\hbar\omega-\Delta_0$) is coupled to the singlet state. \textbf{b,c} show the fraction of double occupancy and singlet states in the Floquet-eigenstate. The shaded areas mark the interaction regimes in which the effective Hamiltonians $H_{\text{eff}}^{(0)}$ (cyan) and $H_{\text{eff}}^{(2)}$ (brown) are valid. 
		\textbf{d,}
		Experimentally measured double occupancy fraction for $t/h=260(30)\:\text{Hz}$ and $\Delta_0/h=660(20)\:\text{Hz}$ as a function of $U$ for a static double well (red) and a two-frequency drive (blue) with $\omega/(2\pi)=2.75\:\text{kHz}$, $2\omega/(2\pi)=5.5\:\text{kHz}$ and $\phi_{\text{r}}=\pi$. The driving strength was ramped up at the final interaction $U$ within $5.454\:\text{ms}$ (15 modulation cycles), followed by an additional modulation for $10.182\:\text{ms}$ (28 cycles) at fixed amplitudes of $K_1=K_2=0.79(1)$. Solid lines are guides to the eye. We observe resonances at $U=l\hbar\omega\pm\Delta_0$ for $l=0,...,4$. For $l>0$, we connect to different Floquet-eigenstates depending on whether we ramp up the drive on the left or right side of the resonance, which results in a peak-shaped resonance feature \cite{sDesbuquois2017}. Measurements for interactions $U/h<4\:\text{kHz}$ were performed with a $\{\uparrow,\rightarrow\}$-pair of atoms, while for $U/h>4\:\text{kHz}$ a $\{\uparrow,\downarrow\}$-pair was used (see Methods). Data points and error bars denote mean and standard deviation of 6 individual measurements. 
	}
	\label{fig:S1}
\end{figure*}

Next, let us consider the full effective Hamiltonian for $U\approx l\hbar\omega$. In general, taking into account Eqs.\;(\ref{Hrot}) and (\ref{teff}), it can be written as 
\begin{eqnarray}
\hat{\widetilde{H}}_{\text{eff}}^{(l)}=&-&\sqrt{2}\left[t_{\text{eff,L}}^{(l)} \hat{H}_{\text{L}}+t_{\text{eff,R}}^{(l)} \hat{H}_{\text{R}} + \text{h.c.} \right]\notag\\
&+& (U-l\hbar\omega) \hat{H}_U + \Delta_0 \hat{H}_{\Delta} \hspace{4mm} \in \mathbb{C}^{4\times4}. \label{Heff4}
\end{eqnarray}
The effective interaction in the near-resonantly system is again given by $U-l\hbar\omega$. For any value of $l$, there are two resonances appearing at $U=l\hbar\omega\pm\Delta_0$, where the singlet state is coupled to either of the two double occupancy states $\left|0,\uparrow\downarrow\right\rangle$ or $\left|\uparrow\downarrow,0\right\rangle$, respectively (see Supp.\;Fig.\;\ref{fig:S1}). In the case where $\hbar\omega,U\gg\Delta_0 \gg t$, these resonances are well separated and we can selectively couple the singlet state only to one of the two double ocupancy states by choosing a suitable driving frequency. If we focus e.g. on the resonance $U=l\hbar\omega+\Delta_0$, we can truncate the Hamiltonian and restrict ourselves to an effective two-level system of the double occupancy state $\left|0,\uparrow\downarrow\right\rangle$ and the singlet state $\left|\text{s}\right\rangle$ (see also Fig.\;\ref{fig:1}b,c). In this basis, the Hamiltonian can be written as
\begin{equation}
\hat{H}_{\text{eff}}^{(l)}=\vec{h}^{(l)}\cdot\vec{\sigma}+\frac{\delta^{(l)}}{2}\mathds{1}^{2\times 2}   \hspace{4mm} \in \mathbb{C}^{2\times2}
\label{Heff2}
\end{equation}
with the vector
\begin{equation}
\vec{h}^{(l)}=\left(-\sqrt{2}|t_{\text{eff}}^{(l)}|\cos(\psi^{(l)}),\sqrt{2}|t_{\text{eff}}^{(l)}|\sin(\psi^{(l)}),\delta^{(l)}/2\right)
\label{HeffVector}
\end{equation}
and $\vec{\sigma}=(\sigma_x,\sigma_y,\sigma_z)$ is the vector of the Pauli spin matrices. The detuning from the resonance is given by $\delta^{(l)}=U-l\hbar\omega-\Delta_0$ and we set $t_{\text{eff}}^{(l)}\equiv t_{\text{eff},\text{R}}^{(l)}$ (the respective expression for the resonance $U=l\hbar\omega-\Delta_0$ can be derived by replacing $\delta^{(l)}=U-l\hbar\omega+\Delta_0$ and $t_{\text{eff}}^{(l)}\equiv t_{\text{eff},\text{L}}^{(l)}$ while taking into account relation (\ref{tpm})). 

The eigenenergies of the effective Hamiltonian (\ref{Heff2}) are given by
\begin{equation}
\epsilon_{1,2}^{(l)}=\left(\delta^{(l)}\pm \sqrt{(\delta^{(l)})^2+8|t_{\text{eff}}^{(l)}|^2} \right)/2.
\end{equation}
Exactly on resonance $\delta^{(l)}=0$, they reduce to  
\begin{equation}
\epsilon_{1,2}^{(l)}=\pm\sqrt{2}|t_{\text{eff}}^{(l)}|
\end{equation}
while the corresponding eigenstates are given by
\begin{equation}
\left|\varphi_{1,2}^{(l)}\right\rangle = \frac{1}{\sqrt{2}}\begin{pmatrix} \mp e^{i\psi^{(l)}} \\
									1 \end{pmatrix}.
\label{eigenstates}
\end{equation}
In a representation on the Bloch sphere, $\left|\varphi_{1,2}^{(l)}\right\rangle$ point (anti-)parallel to the Hamiltonian $\hat{H}_{\text{eff}}^{(l)}$ given by the vector (\ref{HeffVector}) with $\delta^{(l)}=0$.

\subsection{Different tunnelling regimes and $\mathbb{Z}_2$-invariant}
We now analyse the structure of the effective matrix elements given in Eq.\;(\ref{teff}). We will focus on $t_{\text{eff}}^{(0)}$ and $t_{\text{eff}}^{(2)}$, which are the ones that are investigated in the experiments. The leading terms of the series in Eq.\;(\ref{teff}) (keeping Besselfunctions $\mathcal{J}_{\nu}(K_{1,2})$ up to order $\nu=2$) are given by
\begin{eqnarray}
t_{\text{eff}}^{(0)} &=& t \left[\mathcal{J}_0(K_1)\mathcal{J}_0(K_2) \right.\notag\\
			&& \left. -2i \mathcal{J}_2(K_1)\mathcal{J}_1(K_2)\sin(\phi_{\text{r}}) + ... \right] \label{t0}\\
t_{\text{eff}}^{(2)} &=& t e^{-2i\phi_{\text{c}}} \left[\mathcal{J}_2(K_1)\mathcal{J}_0(K_2) \right.\notag\\
			&& \left. + \mathcal{J}_0(K_1)\mathcal{J}_1(K_2)e^{-i\phi_{\text{r}}} + ... \right]. \label{t2}
\end{eqnarray}
We can rewrite them as
\begin{eqnarray}
t_{\text{eff}}^{(0)} &=& t\left[\alpha^{(0)}(K_1,K_2) \right. \notag \\
						&& \left. -2i\beta^{(0)}(K_1,K_2)\sin(\phi_{\text{r}})\right] \label{t0Simple} \\
t_{\text{eff}}^{(2)} &=& t\left[\alpha^{(2)}(K_1,K_2)+\beta^{(2)}(K_1,K_2)e^{-i\phi_{\text{r}}} \right], \label{t2Simple}
\end{eqnarray}
where $\alpha^{(l)}(K_1,K_2)$, $\beta^{(l)}(K_1,K_2)>0$ can be directly identified by comparing the expressions to Eqs.\;(\ref{t0}) and (\ref{t2}) (in the following we will omit the explicit dependence of $\alpha^{(l)}$ and $\beta^{(l)}$ on the driving amplitudes).

We first focus on the expression for $t_{\text{eff}}^{(2)}$. If we represent the tunnelling matrix element in the complex plane for values $\phi_{\text{r}} \in [0,2\pi)$, it describes a circle around the point $(\alpha^{(2)},0)$ with radius $\beta^{(2)}$. Importantly, there are two distinct regimes: For $\beta^{(2)}<\alpha^{(2)}$, the circle is entirely located in the right half of the complex plane for which $\text{Re}\left[t_{\text{eff}}^{(2)}\right]>0$, while for $\beta^{(2)}>\alpha^{(2)}$ it encloses the origin $(0,0)$. This means that in the latter case, for $\delta^{(2)}=0$ the Hamiltonian represented by $\vec{h}^{(2)}$ (see Eq.\;(\ref{HeffVector})) is winding once around the Bloch sphere when changing the relative phase $\phi_{\text{r}}$ from $0$ to $2\pi$. In contrast, for $\beta^{(2)}<\alpha^{(2)}$ the tunnelling phase only takes small values $\psi_{\mathrm{r}}^{(2)}\in (-\pi/2,\pi/2)$ and the Hamiltonian does not go around the entire Bloch sphere. These two regimes are separated by the special point where $\beta^{(2)}=\alpha^{(2)}$, for which $t_{\text{eff}}^{(2)}=0$ at $\phi_{\text{r}}=\pi$. 

\begin{figure}
	\includegraphics[width=89mm]{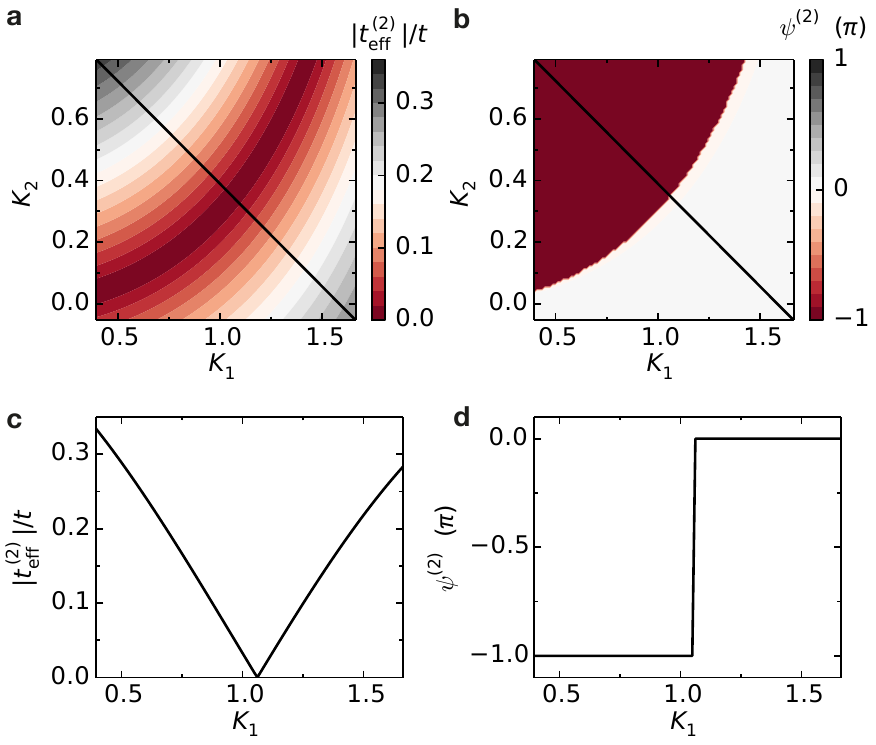}
	\caption{
		\textbf{Theory for the gap closing in the $K_1$-$K_2$ parameter space (compare to Fig.\;\ref{fig:3}).}
		Effective tunnelling matrix element $t_{\text{eff}}^{(2)}\equiv t_{\text{eff,R}}^{(2)}$ according to Eq.\;(\ref{teff}) for $\phi_{\text{c}}=0$ up to terms with $\left|m\right|\leq 4$.
	\textbf{a,b,}
	Absolute value and phase of $t_{\text{eff}}^{(2)}$ as a function of the driving amplitudes $K_1$ and $K_2$ for a relative phase of $\phi_{\text{r}}=\pi$. The absolute value is vanishing along a line (dark red colour in \textbf{a}) which is to lowest order given by $\alpha^{(2)}=\beta^{(2)}$, i.e. $\mathcal{J}_2(K_1)\mathcal{J}_0(K_2)=\mathcal{J}_0(K_1)\mathcal{J}_1(K_2)$. This line separates the parameter space into two different regimes: In the bottom right area (white colour in \textbf{b}), the $\omega$-drive is dominant and the effective tunnelling is positive, while in the top left corner the phase is given by $\pi$ (dark red colour in \textbf{b}). These are the only possible values of $\psi_{\mathrm{r}}^{(2)}$, since for $\phi_{\text{r}}=\pi$ time-reversal symmetry is not broken and the tunnelling has to be real-valued. 
	\textbf{c,d,}
	Absolute value and phase of $t_{\text{eff}}^{(2)}$ along the black cut in \textbf{a} and \textbf{b}, showing the linear dependence of $|t_{\text{eff}}^{(2)}|$ on the driving amplitude around the gap closing and the sudden jump of the tunnelling phase at $K_1=K_{1,\text{crit}}=1.06(1)$. 
	}
	\label{fig:S2}
\end{figure}

More specifically, this transition (in the following referred to as 'gap closing', since at this point $|t_{\text{eff}}^{(2)}|=0$) occurs at $\mathcal{J}_2(K_1)\mathcal{J}_0(K_2)\approx \mathcal{J}_0(K_1)\mathcal{J}_1(K_2)$ and $\phi_{\text{r}}=\pi$. This can be demonstrated by plotting the absolute value of the tunnelling versus $K_1$ and $K_2$, see Supp.\;Fig.\;\ref{fig:S2}a. This gap closing comes together with a sign change of the effective tunnelling, which means that $\psi_{\mathrm{r}}^{(2)}$ jumps by $\pi$ (see Supp.\;Fig.\;\ref{fig:S2}b,d). This can be seen directly in expression (\ref{t2Simple}): for $\phi_{\text{r}}=0$ and $\pi$, the effective tunnelling is always real (this is a consequence from the fact that the waveform is TR symmetric). However, while for $\phi_{\text{r}}=0$, $t_{\text{eff}}^{(2)}$ is always positive, it changes sign for $\phi_{\text{r}}=\pi$ at $\beta^{(2)}=\alpha^{(2)}$. Therefore, it is enough to look at the sign of the tunnelling at $\phi_{\text{r}}=\pi$ to determine in which regime we are: if $t_{\text{eff}}^{(2)}<0$ at this point, the Hamiltonian is wrapping around the Bloch sphere when changing $\phi_{\text{r}}$ from $0$ to $2\pi$, while for $t_{\text{eff}}^{(2)}>0$ it does not. 

\begin{figure}
	\includegraphics[width=89mm]{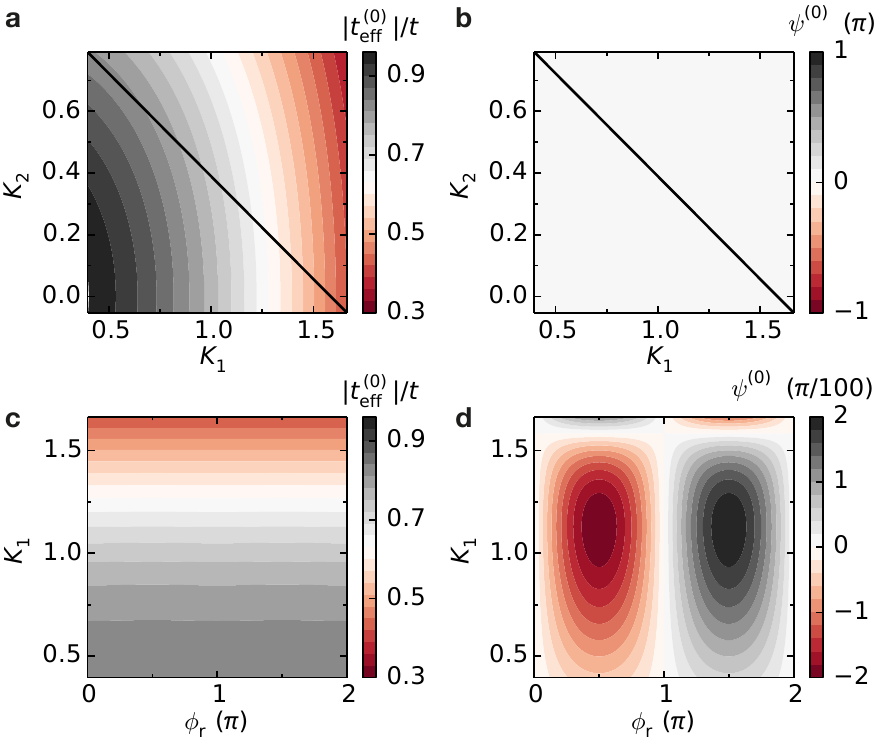}
	\caption{
		\textbf{Theory for the single-particle tunnelling.}
		Effective tunnelling matrix element $t_{\text{eff}}^{(0)}$ according to Eq.\;(\ref{teff}) for $\phi_{\text{c}}=0$ up to terms with $\left|m\right|\leq 4$. 
	\textbf{a,b,}
	Absolute value and phase of $t_{\text{eff}}^{(0)}$ as a function of the driving amplitudes $K_1$ and $K_2$ for a relative phase of $\phi_{\text{r}}=\pi$. The gap does not close in this regime and the phase is constant $\psi^{(0)}=0$. 
	\textbf{c,d,}
	Dependence of the absolute value and phase of $t_{\text{eff}}^{(0)}$ on the relative phase $\phi_{\text{r}}$ and the driving amplitudes. The parametrisation of $K_2$ is chosen along the black line in \textbf{a} and \textbf{b}. While $|t_{\text{eff}}^{(0)}|$ is varying as a function of the driving amplitude, the tunnelling phase is negligable with $\psi^{(0)}<\pi/50$.  
	}
	\label{fig:S3}
\end{figure}

Finally, let us look at the expression for $t_{\text{eff}}^{(0)}$ in Eq.\;(\ref{t0Simple}). In this case, $t_{\text{eff}}^{(0)}(\phi_{\text{r}}=0)=t_{\text{eff}}^{(0)}(\phi_{\text{r}}=\pi)\in\mathbb{R}$, which also follows in general from the relations (\ref{teffConj}) and (\ref{teff0Pi}). This means that for $l=0$, the effective Hamiltonian is forced to point in the same direction at the two TR-symmetric points. In order to characterise the properties of the effective tunneling when changing the relative driving phase from $0$ to $2\pi$, we can define the $\mathbb{Z}_2$ invariant
\begin{eqnarray}
\mathcal{Z}^{(l)}&=&\exp\left[i\left(\psi^{(l)}({\phi_{\text{r}}=\pi})-\psi^{(l)}({\phi_{\text{r}}=0})\right)\right] \\
&=& \text{sgn}\left(t_{\text{eff}}^{(l)}({\phi_{\text{r}}=\pi}) \right) \cdot \text{sgn}\left(t_{\text{eff}}^{(l)}({\phi_{\text{r}}=0}) \right) \\
&\in& \left\{-1,1\right\} \notag
\end{eqnarray}
with the sign function $\text{sgn}(x)$. In particular, $\mathcal{Z}^{(0)}=1$ which means that there are no distinct tunnelling regimes for the single-particle hopping $t_{\text{eff}}^{(0)}$ (see Supp.\;Fig.\;\ref{fig:S3}a,b). In contrast, $\mathcal{Z}^{(l)}$ changes sign for $l\neq 0$ at the critical point $\beta^{(l)}=\alpha^{(l)}$. 

\subsection{Dirac point structure of the effective tunnelling}
\begin{figure}[htb]
	\includegraphics[width=89mm]{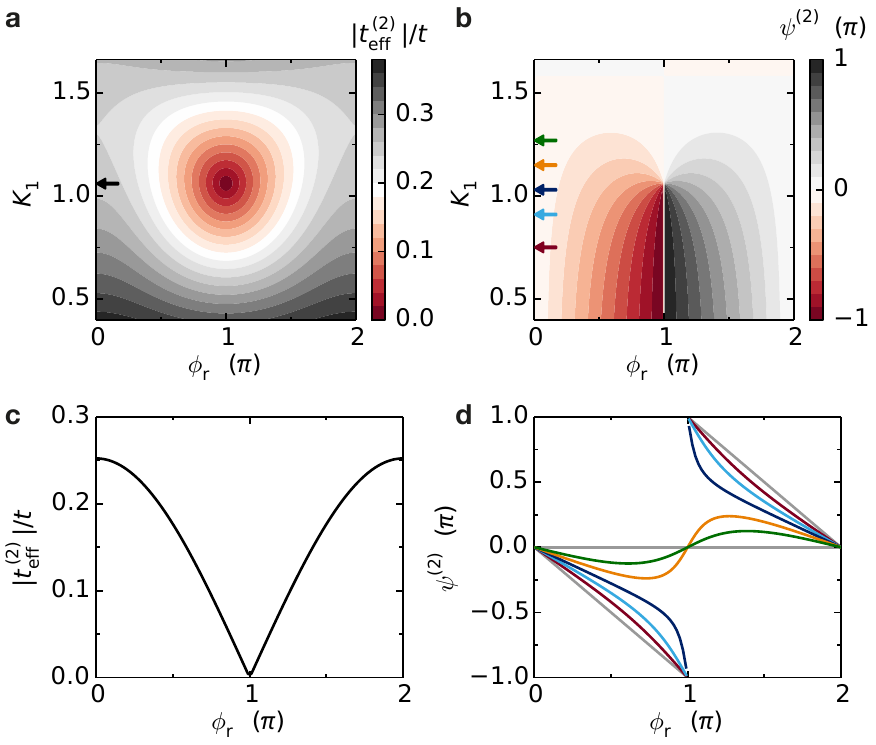}
	\caption{
		\textbf{Theory for the Dirac point and phase vortex (compare to Fig.\;\ref{fig:4}).}
		Effective tunnelling matrix element $t_{\text{eff}}^{(2)}\equiv t_{\text{eff,R}}^{(2)}$ according to Eq.\;(\ref{teff}) for $\phi_{\text{c}}=0$ up to terms with $\left|m\right|\leq 4$.
	\textbf{a,b,}
	Dependence of the absolute value and phase of $t_{\text{eff}}^{(2)}$ on the relative phase $\phi_{\text{r}}$ and the driving amplitudes. The parametrisation of $K_2$ is chosen along the blue line in Supp. Figs.\;\ref{fig:S3} and \ref{fig:S2}. The gap is closing at a singular point in parameter space where $K_1=K_{1,\text{crit}}=1.06(1)$ (black arrow in \textbf{a}) and $\phi_{\text{r}}=\pi$. Away from this point, $|t_{\text{eff}}^{(2)}|$ is increasing linearly as a function of $K_1$ and $\phi_{\text{r}}$ (see also \textbf{c} and Supp.\;Fig.\;\ref{fig:S2}c). The tunnelling phase has a vortex structure close to the Dirac point and is increasing by $2\pi$ when moving clockwise around the singularity. Arrows indicate the values of $K_1$ for the cuts shown in \textbf{c} and \textbf{d}.
	\textbf{c,}
	Cut through \textbf{a} at $K_1=K_{1,\text{crit}}=1.06(1)$ showing the linear closing of the gap as a function of the relative phase around $\phi_{\text{r}}=\pi$.
	\textbf{d,}
	Cuts through \textbf{b} at fixed driving amplitudes $K_1=$0.75, 0.91, 1.03, 1.15 and 1.27 indicated by the arrows in \textbf{b}. The gray lines correspond to the limiting cases $K_2=0$ (with $\psi_{\mathrm{r}}^{(2)}=0$) and $K_1=0$ (with $\psi_{\mathrm{r}}^{(2)}=-\phi_{\text{r}}$), respectively. For large amplitudes $K_1>K_{1,\text{crit}}$, the tunnelling phase is only taking small values between $-\pi/2$ and $\pi/2$. In contrast, for $K_1<K_{1,\text{crit}}$ we observe a running phase and $\psi_{\mathrm{r}}^{(2)}$ is changing from $0$ through $\mp\pi$ back to $0$. In this case, the Hamiltonian is winding once around the Bloch sphere when sweeping the relative phase from $0$ to $2\pi$. Note that at the time-reversal symmetric points where $\phi_{\text{r}}=0$ and $\pi$, the tunnelling phase is forced to be either $0$ or $\pi$ since $t_{\text{eff}}^{(2)}$ is real.
	}
	\label{fig:S4}
\end{figure}

We now want to investigate the structure of the effective Hamiltonian $\hat{H}_{\text{eff}}^{(2)}$ around the special point where $\alpha^{(2)}=\beta^{(2)}$ and $\phi_{\text{r}}=\pi$. For small deviations of the phase $\tilde{\phi}_{\text{r}}=\phi_{\text{r}}-\pi$, the effective tunnelling is given by 
\begin{equation}
t_{\text{eff}}^{(2)}\approx \alpha^{(2)}-\beta^{(2)}+i\beta^{(2)}\tilde{\phi}_{\text{r}}
\end{equation}
with an absolute value
\begin{equation}
|t_{\text{eff}}^{(2)}| = \sqrt{\left[\alpha^{(2)}-\beta^{(2)}\right]^2+\left[\beta^{(2)}\tilde{\phi}_{\text{r}}\right]^2}.
\end{equation}
The gap is therefore linearly increasing in the parameters $\alpha^{(2)}-\beta^{(2)}$ and $\tilde{\phi}_{\text{r}}$ around the point where $t_{\text{eff}}^{(2)}=0$. We can additionally expand $\alpha^{(2)}=\mathcal{J}_2(K_1)\mathcal{J}_0(K_2)$ and $\beta^{(2)}=\mathcal{J}_0(K_1)\mathcal{J}_1(K_2)$ in the experimental parameters $K_1$ and $K_2$. For this, we focus on a specific parametrization of $K_1$ and $K_2$ shown in Supp. Figs.\;\ref{fig:S2} and \ref{fig:S3}a,b (here $K_2=2/3(m-K_1)$ with $m=1.58(1)$). For this choice, the gap is closing at a critical amplitude $K_{1\text{crit}}=1.06(1)$ for $\phi_{\text{r}}=\pi$. Expanding the effective tunnelling up to linear order in $K_1$ and $\phi_{\text{r}}$ around this point gives
\begin{equation}
\frac{t_{\text{Dirac}}}{t}=\frac{c_K}{\sqrt{2}}(K_1-K_{1,\text{crit}})+i \frac{c_{\phi}}{\sqrt{2}}(\phi_{\text{r}}-\pi),
\end{equation}
where the numerical factors $K_{1,\text{crit}}$, $c_K=0.537(1)$ and $c_{\phi}=\beta^{(2)}(K_{1\text{crit}},K_{2\text{crit}})=0.123(1)$ depend on the $K_1$-$K_2$ parametrization. Around the gap closing point, the Hamiltonian in Eq.\;(\ref{Heff2}) can therefore be written as
\begin{equation}
\hat{H}_{\text{Dirac}}=-t c_K \tilde{K}_1 \sigma_x+t c_{\phi} \tilde{\phi}_{\text{r}} \sigma_y
\end{equation}
for $\delta^{(2)}=0$ with $\tilde{K}_1=K_1-K_{1,\text{crit}}$ and $\tilde{\phi}_{\text{r}}=\phi_{\text{r}}-\pi$. 
This is a Dirac Hamiltonian in the experimental parameters of driving amplitudes and relative phase, which only affects the density-assisted tunnelling processes, while the single-particle hopping remains trivial (see Supp.\;Fig.\;\ref{fig:S3}c,d). 
In particular, the absolute value of the tunnelling amplitude increases linearly away from the Dirac point with 
\begin{equation}
\left|t_{\text{Dirac}}\right|=\frac{t}{\sqrt{2}} \sqrt{(c_K \tilde{K}_1)^2+(c_{\phi} \tilde{\phi}_{\text{r}})^2},
\end{equation}
while the phase has a vortex structure around the singularity with 
\begin{equation}
\tan{(\psi_{\text{Dirac}})}=c_{\phi} \tilde{\phi}_{\text{r}}/(c_K \tilde{K}_1)
\end{equation}
(see Supp.\;Fig.\;\ref{fig:S4}). We can see again that for values $K_1>K_{1,\text{crit}}$, $\psi_{\text{Dirac}}$ only takes small positive and negative values, while for small values of $K_1$ it is running from $0$ through $\mp\pi$ back to $0$ (see Supp.\;Fig.\;\ref{fig:S4}d). These regions correspond to the distinct regimes discussed above, which are characterised by the $\mathbb{Z}_2$ invariant. 

Finally we want to mention that there also exist Dirac points for the case $l=0$. However, due to relation (\ref{teff0Pi}), they always come in pairs at two relative modulation phases $\phi_{\text{r}}$ and $\phi_{\text{r}}+\pi$. For example, to lowest order two Dirac points appear for $\alpha^{(0)}=0$ at $\phi_{\text{r}}=0$ and $\phi_{\text{r}}=\pi$ (see Eq.\;(\ref{t0Simple})). In general, due to the complex conjugation in Eq.\;(\ref{teff0Pi}), the winding sense for the two Dirac points is always opposite. Therefore, when sweeping $\phi_{\text{r}}$ from $0$ to $2\pi$, their contributions to the tunnelling phase add up and cancel each other, such that the Hamiltonian $\vec{h}^{(0)}$ cannot wrap around the Bloch sphere. Furthermore, for $l>0$ one can also end up with a pair of Dirac points with opposite winding at $\phi_{\text{r}}=0$ and $\phi_{\text{r}}=\pi$ for (almost) the same values of $K_1$ and $K_2$. This is for example the case for $\beta^{(2)}=0$, such that the terms with $m$ odd in the expansion (\ref{teff}) disappear to lowest order. The Dirac points then appear when $\alpha^{(2)}=\gamma^{(2)}$ for $\phi_{\text{r}}=0$ and $\pi$, where $\gamma^{(2)}$ is the prefactor of the term $\exp{(-2i\phi_{\text{r}})}$. Corrections only arise from higher order terms with $m$ odd. In addition, the Dirac points are not bound to appear at $\phi_{\text{r}}=0$ and $\pi$, since in general they only have to fulfill the relations (\ref{tpm}) and (\ref{teffConj}). For increasing driving amplitudes $K_1$ and $K_2$, a pair of Dirac points with the {\itshape same} winding sense can for example exist at two arbitrary phases $\pi+\phi_{\text{r}}$ and $\pi-\phi_{\text{r}}$ ($\phi_{\text{r}}\in (0,\pi)$) with the general relation $t_{\text{eff}}^{(l)}(\pi+\phi_{\text{r}})=\left[t_{\text{eff}}^{(l)}(\pi-\phi_{\text{r}})\right]^*$. When changing the driving amplitudes, the creation and annihilation of vortex-antivortex pairs of Dirac points can be observed. 

\section{Part C: Protocols for gap and phase measurements}
\subsection{Measurement of the absolute value of the tunneling matrix element}
In the experiment, we measure the absolute value of the effective tunnelling $|t_{\text{eff}}^{(2)}|$ on the resonance $U=2\hbar\omega+\Delta_0$ by ramping $\delta^{(2)}$ in a Landau-Zener-type experiment from negative to positive values across the resonance. We measure how much of the population follows adiabatically the eigenstate $\left|\varphi_{1}^{(2)}\right\rangle$ (see Eq.\;(\ref{eigenstates}) and Fig.\;\ref{fig:2}a), which means that the state is converted from a singlet to a double occupancy state. To estimate the amount of adiabatic transfer, we can use the Landau-Zener formula for the probability of staying in the ground state
\begin{equation}
P_{\text{adiab}}=1-e^{-\Gamma^2}
\end{equation}
with
\begin{equation}
\Gamma = 2\sqrt{2}\pi\left|t_{\text{eff}}^{(2)}/h\right| \left[\frac{\text{d}(\delta^{(2)}/h)}{\text{d}\tau}\right]^{-1/2}.
\end{equation}
In our concrete case, we use a linear ramp of the detuning over a span of 2.5(1)\;kHz within 20\;ms, for which we find $\Gamma=|t_{\text{eff}}^{(2)}|/(\kappa\cdot t)$ with $\kappa=0.15(2)$. This means that the characteristic energy scale that we can resolve with the measurement is on the order of $\kappa\cdot t\approx 40\:\text{Hz}$. We confirm this estimate in the experiment by measuring the gap size for a single frequency drive (see Fig.\;\ref{fig:2}b).

When looking at the results of the gap measurement in Fig.\;\ref{fig:3}a, we see that the double occupancy fraction almost goes to zero for small values of both $K_1$ and $K_2$ along the line where $|t_{\text{eff}}^{(2)}|=0$, while the minimum is less pronounced if both amplitudes are high. One reason can be that in this region, the absolute value of the tunneling amplitude is very sensitive on the parameters $K_1$, $K_2$ and $\phi_{\text{r}}$. Hence, the reduced contrast of the gap measurement could result from experimental shot-to-shot fluctuations of the modulation parameters, which increases the measured average gap size. On the other hand, the reduced contrast could result from the gap measurement sequence itself. 

\begin{figure*}
	\includegraphics[width=178mm]{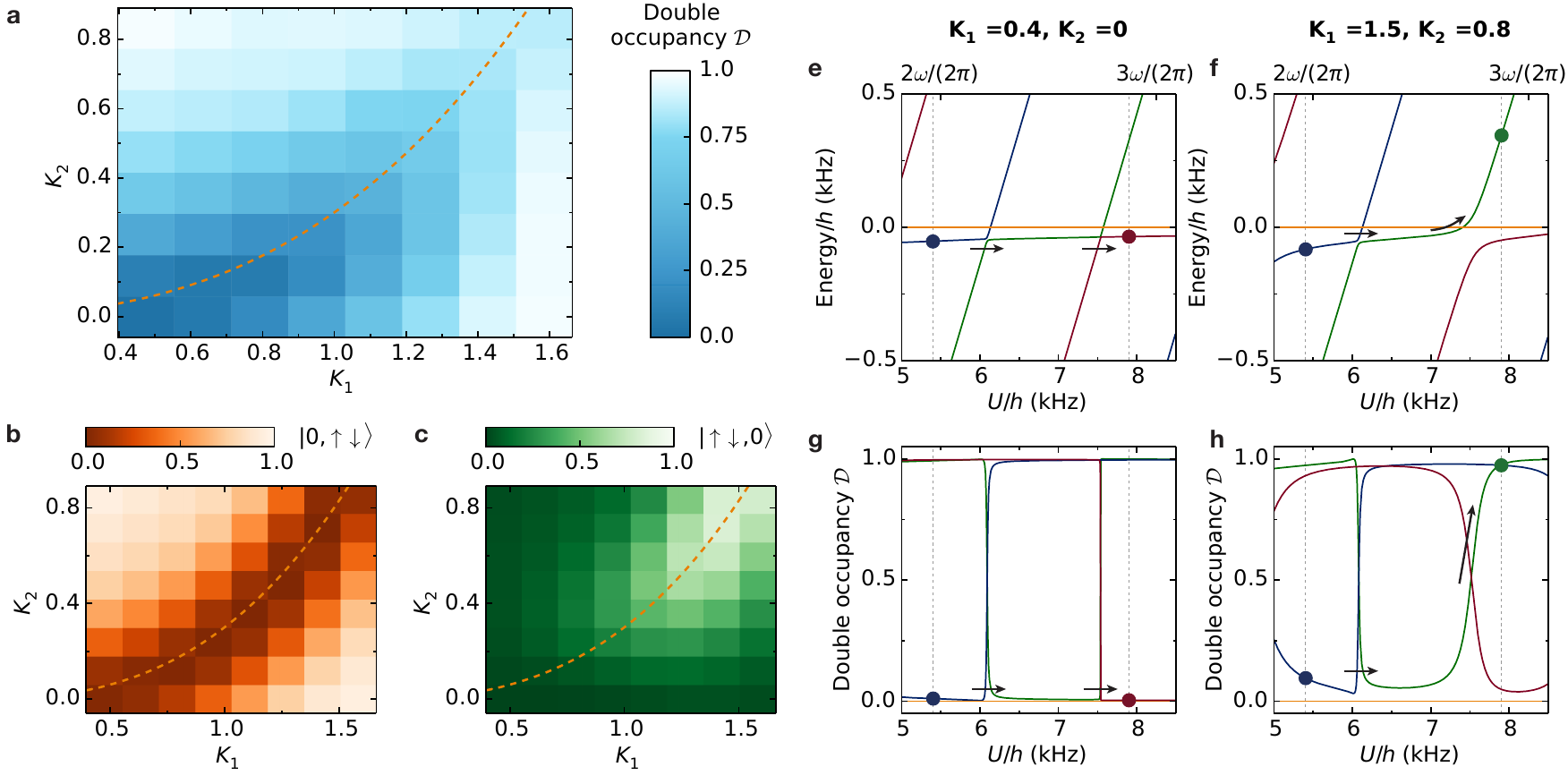}
	\caption{
		\textbf{Numerical simulation of the gap measurement in the $K_1$-$K_2$ parameter space.}
		\textbf{a-c,}
		Numerically calculated double occupancy populations when ramping across the $U=2\hbar\omega+\Delta_0$ resonance as a function of the modulation amplitudes $K_1$ and $K_2$ for the same parameters as in the experiment in Fig.\;\ref{fig:3}a (for details see Methods). The orange dashed line marks the theoretical value for which $|t_{\text{eff}}^{(2)}|=0$ according to Eq.\;(\ref{teff}). 
		The total double occupancy fraction in the final state $\mathcal{D}=\left|\left\langle 0,\uparrow\downarrow\right.\left|\varphi_{\text{fin}}\right\rangle\right|^2 + \left|\left\langle\uparrow\downarrow,0\right.\left|\varphi_{\text{fin}}\right\rangle\right|^2$ is increasing in the regime where the gap is closing for high value of $K_1$ and $K_2$ (\textbf{a}). This is similar to the experimental results in Fig.\;\ref{fig:3}a, where the value of $\mathcal{D}$ is additionally limited by the fraction of doubly occupied dimers $\mathcal{D}_{\text{max}}=0.56(2)$. 
		The projection onto the state $\left|\left\langle 0,\uparrow\downarrow\right.\left|\varphi_{\text{fin}}\right\rangle\right|^2$ is always negligible along the gap closing line (\textbf{b}), while the population of the other double occupancy state $\left|\left\langle\uparrow\downarrow,0\right.\left|\varphi_{\text{fin}}\right\rangle\right|^2$ increases to finite values at high values of $K_1$ and $K_2$ (\textbf{c}). This explains the higher value of the total double occupancy in \textbf{a}.
		\textbf{e-h,}
		Quasi-energy spectrum and double occupancy content of the Floquet-eigenstates as a function of $U$ for different choices of the driving amplitudes located near the gap closing line in \textbf{a-c}. The vertical dashed lines mark the initial and final values of the interaction sweep used for the gap measurement. Circles and arrows mark the initial and final states and the evolution during the gap measurement protocol, respectively. 
		For $K_1=0.4$ and $K_2=0$ (\textbf{e}, \textbf{g}), the gap at $U=3\hbar\omega-\Delta_0$ is even smaller than the one at $U=2\hbar\omega+\Delta_0$. Therefore, after diabatically crossing the first resonance, the atoms will also remain in a singlet state upon passing the second resonance.
		The situation is different for $K_1=1.5$ and $K_2=0.8$ (\textbf{f}, \textbf{h}), where the gap at $U=3\hbar\omega-\Delta_0$ is much larger than the one at $U=2\hbar\omega+\Delta_0$. As a result, even though the atoms remain in a singlet state when crossing the first resonance, they will adiabatically follow the Floquet eigenstate at the second resonance and therefore be converted to the $\left|\uparrow\downarrow,0\right\rangle$ state (see \textbf{c}). Since the measurement cannot distinguish between the two double occupancy states, the total fraction of $\mathcal{D}$ in the experiment increases in this regime even though $|t_{\text{eff}}^{(2)}|=0$ (see \textbf{a}).
	}
	\label{fig:S5}
\end{figure*}

In order to investigate this and to gain a better understanding of the results in Fig.\;\ref{fig:3}a, we perform a full numerical simulation of the gap measurement. As shown in Supp.\;Fig.\;\ref{fig:S5}a, the result is very similar to the observations in the experiment. In particular, the contrast of the total double occupancy fraction also decreases when going to larger values of the driving amplitudes. Interestingly, when looking independently at the two double occupancy states $\left|0,\uparrow\downarrow\right\rangle$ and $\left|\uparrow\downarrow,0\right\rangle$ (Supp.\;Fig.\;\ref{fig:S5}b,c), we see that the population of the desired state $\left|0,\uparrow\downarrow\right\rangle$ is indeed always close to zero along the line where $|t_{\text{eff}}^{(2)}|=0$. However, for large values of $K_1$ and $K_2$, the fraction of $\left|\uparrow\downarrow,0\right\rangle$ also takes finite values, which means that we left the two level system spanned by the singlet state and $\left|0,\uparrow\downarrow\right\rangle$. As a result, the total double occupancy fraction $\mathcal{D}$ also becomes finite in this regime. 

\begin{figure}
	\includegraphics[width=89mm]{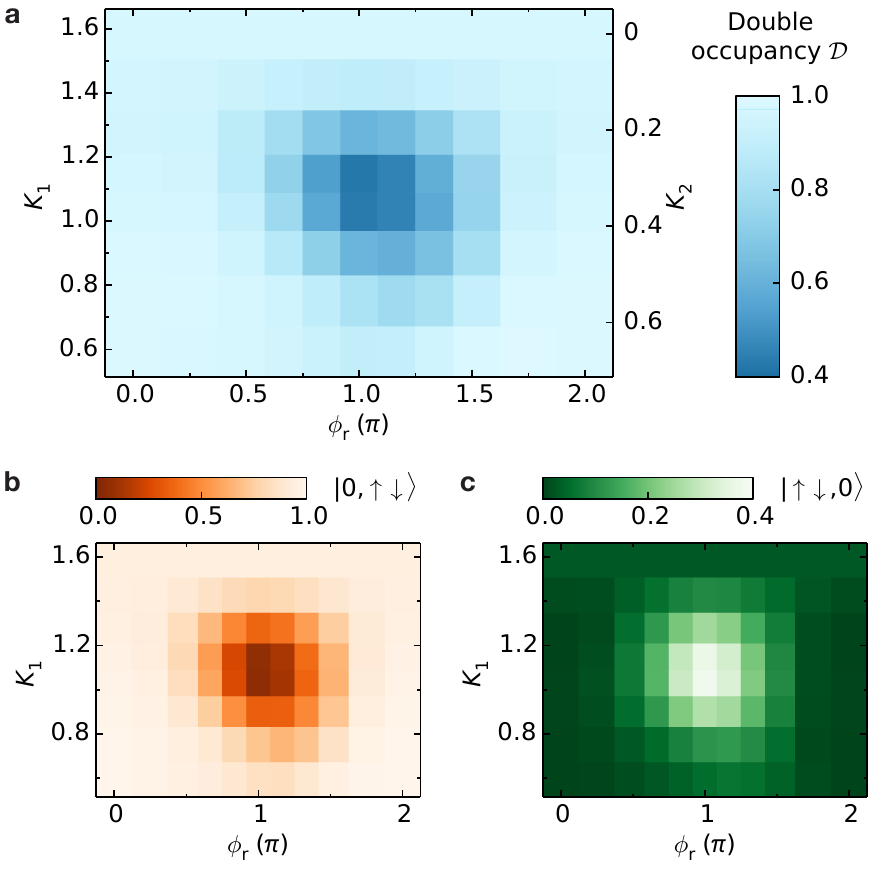}
	\caption{
		\textbf{Numerical simulation of the Dirac point gap measurement.}
		Numerically calculated double occupancy populations when ramping across the $U=2\hbar\omega+\Delta_0$ resonance as a function of the relative phase $\phi_{\text{r}}$ and the driving amplitudes $K_1$-$K_2$ for the same parameters as in the experiment in Fig.\;\ref{fig:4}a (for details see Methods). The amplitudes are parametrized along the black line in Fig.\;\ref{fig:3}a. 
		\textbf{a,}
		The total double occupancy fraction in the final state $\mathcal{D}=\left|\left\langle 0,\uparrow\downarrow\right.\left|\varphi_{\text{fin}}\right\rangle\right|^2 + \left|\left\langle\uparrow\downarrow,0\right.\left|\varphi_{\text{fin}}\right\rangle\right|^2$ is not vanishing at the Dirac point similar to the observations in the experiment, where the value of $\mathcal{D}$ is additionally limited by the fraction of doubly occupied dimers $\mathcal{D}_{\text{max}}=0.56(2)$. 
		\textbf{b,c,}
		Projections onto the individual double occupancy states $\left|\left\langle 0,\uparrow\downarrow\right.\left|\varphi_{\text{fin}}\right\rangle\right|^2$ and $\left|\left\langle\uparrow\downarrow,0\right.\left|\varphi_{\text{fin}}\right\rangle\right|^2$, respectively. Just as in Supp.\;Fig.\;\ref{fig:S5}, the population of the state $\left|0,\uparrow\downarrow\right\rangle$ is vanishing at the gap closing, while the fraction of $\left|\uparrow\downarrow,0\right\rangle$ increases to finite values at the Dirac point. This phenomenon, which is explained in Supp.\;Fig.\;\ref{fig:S5}e-h, is responsible for the higher value of the total double occupancy at the gap closing point in \textbf{a}.
	}
	\label{fig:S6}
\end{figure}

The reason for this behavior can be understood by looking at the spectrum of the driven double well (see Supp.\;Fig.\;\ref{fig:S5}e,f). For our choice of the final interaction $U/h=7.9(1)\:\text{kHz}$ at the end of the sweep, we do not only cross the resonance at $U=2\hbar\omega+\Delta_0$, but also the next one at $U=3\hbar\omega-\Delta_0$. At the latter resonance, the singlet state is coupled to $\left|\uparrow\downarrow,0\right\rangle$. This means that if we adiabatically convert $\left|\text{s}\right\rangle$ to $\left|0,\uparrow\downarrow\right\rangle$ at the first resonance, the state should not be affected at $U=3\hbar\omega-\Delta_0$. However, if the first gap is vanishing for $|t_{\text{eff}}^{(2)}|=0$, we diabatically cross the resonance at $U=2\hbar\omega+\Delta_0$ and stay in the singlet state. In this case, the second resonance at $U=3\hbar\omega-\Delta_0$ becomes relevant. For driving amplitudes chosen in the bottom left corner of Supp.\;Fig.\;\ref{fig:S5}a ($K_1=0.4$ and $K_2=0$), the second gap is even smaller than the first one and we still stay in the singlet state (see Supp.\;Fig.\;\ref{fig:S5}e,g). In contrast, for $K_1=1.5$ and $K_2=0.8$ in the top right corner of Supp.\;Fig.\;\ref{fig:S5}a, the gap at $U=3\hbar\omega-\Delta_0$ is large such that we adiabatically convert $\left|\text{s}\right\rangle$ to $\left|\uparrow\downarrow,0\right\rangle$. Since the experimental measurement cannot distinguish between the two double occupancy states, we obtain a finite value of $\mathcal{D}$ as shown in Supp.\;Fig.\;\ref{fig:S5}a. The same phenomenon also reduces the contrast of the gap measurement around the Dirac point (see Supp.\;Fig.\;\ref{fig:S6}). 

\subsection{Peierls phase measurement}
To measure the phase $\psi^{(2)}$ of the effective tunnelling matrix element $t_{\text{eff}}^{(2)}$ in the experiment, we adiabatically prepare the eigenstate of the resonant Hamiltonian $\hat{H}_{\text{eff}}^{(2)}$ for $\delta^{(2)}=0$ given by
\begin{equation}
\left|\varphi_{1}^{(2)}\right\rangle = \frac{1}{\sqrt{2}}\begin{pmatrix} - e^{i\psi^{(2)}} \\
									1 \end{pmatrix}.
\end{equation}
by slowly ramping the interactions on the resonance $U=2\hbar\omega+\Delta_0$ (see Eq.\;(\ref{eigenstates})). Afterwards, we project the state onto the Hamiltonian $\hat{H}_{\text{eff}}^{(0)}$ with $\delta^{(0)}=0$ and let it evolve (see Fig.\;\ref{fig:2}c). If we assume for simplicity that $\psi^{(0)}=0$ (see Supp.\;Fig.\;\ref{fig:S3}), we have
\begin{equation}
\hat{H}_{\text{eff}}^{(0)} = -\sqrt{2}|t_{\text{eff}}^{(0)}|\sigma_x,
\end{equation}
see Eq.\;(\ref{Heff2}). Hence, the double occupancy after an evolution time $\tau$ will be given by
\begin{eqnarray}
\mathcal{D}(\tau)&=&\left|\left\langle 0,\uparrow\downarrow\left|\exp{\left[{-\frac{i}{\hbar}\hat{H}_{\text{eff}}^{(0)}\tau}\right]}\right|\varphi_{1}^{(2)}\right\rangle\right|^2 \\
&=& \frac{1}{2} \left[1+\sin{\left(\psi^{(2)}\right)}\sin{\left(2\sqrt{2}|t_{\text{eff}}^{(0)}|\tau/\hbar\right)} \right], 
\end{eqnarray}
while the singlet fraction is 
\begin{eqnarray}
\mathcal{S}(\tau)&=&\left|\left\langle \text{s}\left|\exp{\left[{-\frac{i}{\hbar}\hat{H}_{\text{eff}}^{(0)}\tau}\right]}\right|\varphi_{1}^{(2)}\right\rangle\right|^2 \\
&=& \frac{1}{2} \left[1-\sin{\left(\psi^{(2)}\right)}\sin{\left(2\sqrt{2}|t_{\text{eff}}^{(0)}|\tau/\hbar\right)} \right]. 
\end{eqnarray}
This corresponds to coherent oscillations of $\mathcal{D}$ and $\mathcal{S}$ with a frequency of $2\sqrt{2}|t_{\text{eff}}^{(0)}|/h$, an amplitude $\sin{\left(\psi^{(2)}\right)}$ and a relative phase shift of $\pi$. If we fix the time to $\tau=h/(8\sqrt{2}|t_{\text{eff}}^{(0)}|)$ where the state vector rotated by an angle of $\pi/2$ around $\hat{H}_{\text{eff}}^{(0)}$, we see the Ramsey fringes
\begin{equation}
\mathcal{D}(\psi^{(2)}),\mathcal{S}(\psi^{(2)})=\left[1\pm\sin{\left(\psi^{(2)}\right)}\right]/2.
\end{equation}
In the experiment, the actual phase that we measure between the singlet and double occupancy states is given by
\begin{equation}
\psi^{(2)} = -2\omega\tau-2\phi_{\text{c}}+\psi_{\text{r}}^{(2)}. 
\end{equation}
Here, $\psi_{\text{r}}^{(2)}$ only includes the non-trivial part of the Peierls phase of the tunneling matrix element given in Eq.\;(\ref{teff}) without the common phase. The contribution $-2\omega\tau$ is a dynamical phase which appears when going back from the rotating frame to the lab frame via the transformation (\ref{RotFrameTransf}). In order to directly measure the phase of the tunnelling $\psi_{\text{r}}^{(2)} \equiv \psi_{\text{r}}^{(2)}(K_1,K_2,\phi_{\text{r}})$, we scan the absolute phase $\phi_{\text{c}}$ from $0$ to $\pi$ for a fixed time $\tau$. Afterwards, we fit a sine to the resulting fringe and extract $\psi_{\text{r}}^{(2)}$ as the phase shift (see Fig.\;\ref{fig:2}d and Methods).

\subsection{Experimental investigation of the relation between $t_{\text{eff,R}}^{(2)}$ and $t_{\text{eff,L}}^{(2)}$}
Finally we investigate the relation between the two density-assisted tunneling matrix elements $t_{\text{eff,R}}^{(2)}$ and $t_{\text{eff,L}}^{(2)}$, which are associated to the creation and annihilation of a double occupancy via a hopping process to the right (R) and left (L), respectively (see the effective Hamiltonian in Eq.\;(\ref{HeffMany})). In other words we measure the remaining matrix elements that appear in the tunnelling amplitude and gauge field operators (see Eqs.\;(\ref{TunnelOp}) and (\ref{GaugeOp})). 

To this end, we project the system again on a double well, for which the effective Hamiltonian is given by Eq.\;(\ref{Heff4}). In order to map out the matrix element $t_{\text{eff,L}}^{(2)}$, we have to adjust the interactions to be around the resonance $U=2\hbar\omega-\Delta_0$ (instead of $U=2\hbar\omega+\Delta_0$ for the measurement of $t_{\text{eff,R}}^{(2)}$), where the system is again described by an effective two-level system according to Eq.\;(\ref{Heff2}) with $t_{\text{eff}}^{(2)}\equiv t_{\text{eff,L}}^{(2)}$. The easiest way to achieve this is to change the sign of the static site offset $\Delta_0 \rightarrow -\Delta_0$ while keeping all other parameters (such as the static tunneling amplitude and interactions) fixed. We can then measure both the amplitude and Peierls phase of the tunneling matrix element using the same measurement schemes as before. 

\begin{figure}
	\includegraphics[width=89mm]{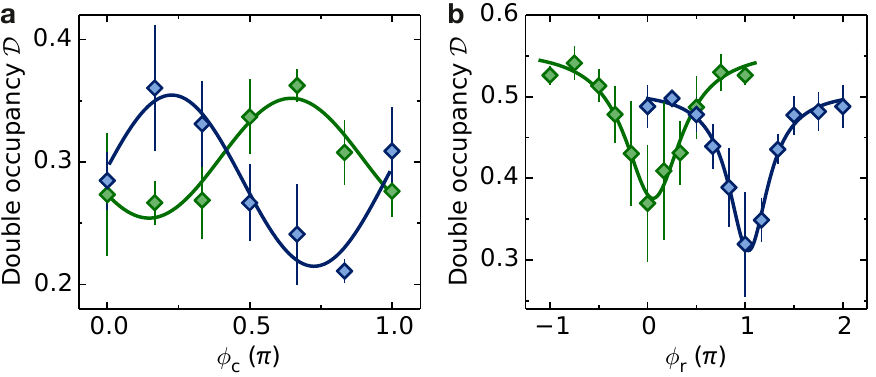}
	\caption{
		\textbf{Change of the effective tunnelling for $\Delta_0 \rightarrow -\Delta_0$.}
		Changing the sign of the static site offset from $\Delta_0/h=660(20)\:\text{Hz}$ (blue) to $\Delta_0/h=-750(20)\:\text{Hz}$ (green) at a fixed interaction strength results in a coupling of the singlet state to $\left|\uparrow\downarrow,0\right\rangle$ instead of $\left|0,\uparrow\downarrow\right\rangle$. With this, we can study the effective tunnelling associated with the creation of a double occupancy on the left site $t_{\text{eff,L}}^{(2)}$ (for $\Delta_0<0$) and compare it to the density-assisted hopping to the right site $t_{\text{eff,R}}^{(2)}$ (for $\Delta_0>0$). 
		\textbf{a,}
		Ramsey fringes of the double occupancy as a function of $\phi_{\text{c}}$ probing the phase of $t_{\text{eff,R/L}}^{(2)}$ for $K_1=0$, $K_2=0.79(1)$ (compare to Fig.\;\ref{fig:2}). As expected, the fringes have a relative phase shift of $\pi$. Solid lines are sinusoidal fits to the data, from which we extract the phase of the Ramsey fringes to be $\psi_{\text{r}}^{(2)}=-0.05^{+0.05}_{-0.06}\pi$ ($\Delta_0>0$) and $\psi_{\text{r}}^{(2)}=0.79^{+0.06}_{-0.07}\pi$ ($\Delta_0<0$), respectively. 
		\textbf{b,}
		Double occupancy fraction when ramping across the $U=2\hbar\omega+\Delta_0$ resonance as a function of the relative phase $\phi_{\text{r}}$ for $K_1=1.03(1)$, $K_2=0.370(5)$ (compare to Fig.\;\ref{fig:4}b). From Lorentzian fits to the data (solid lines), we extract that the gap is closing at a relative phase of $\phi_{\text{r}}=0.06(4)\pi$ ($\Delta_0<0$) instead of $\phi_{\text{r}}=1.03(6)\pi$ ($\Delta_0>0$). This means that the Dirac point is appearing at a different point in the parameter space of the driving parameters $\phi_{\text{r}}$, $K_1$ and $K_2$. 
		Data points and error bars denote mean and standard deviation of 5 individual measurements. 
	}
	\label{fig:S7}
\end{figure}

In Eq.\;(\ref{tpm}) we derived the relation 
\begin{equation}
t_{\text{eff,L}}^{(l)}(\phi_{\text{c}},\phi_{\text{r}}) = t_{\text{eff,R}}^{(l)}(\phi_{\text{c}}+\pi,\phi_{\text{r}}+\pi)
\label{tpmAgain}
\end{equation}
between the two tunnel couplings. In order to investigate the dependence on the common driving phase $\phi_{\text{c}}$, we measure Ramsey fringes for both positive and negative values of $\Delta_0$ for a single frequency drive with $K_1=0$ (see Supp.\;Fig.\;\ref{fig:S7}a). In this case, the matrix elements are given by (see Eq.\;(\ref{teff}))
\begin{equation}
t_{\text{eff,R/L}}^{(2)}=\pm \mathcal{J}_1(K_2),
\end{equation}
which means that the effective tunnelling flips sign upon changing the direction of the density-assisted hopping process. This is a consequence of the reflection property of the Bessel function $\mathcal{J}_1(-K_2)=-\mathcal{J}_1(K_2)$. In the measurement, the Ramsey fringes have a relative phase shift of $\pi$, which confirms the sign change and hence the relations above. 

Next, we investigate the influence of the relative modulation phase $\phi_{\text{r}}$. To this end, we map out the gap closing at the Dirac point near the critical value of the driving amplitude $K_{1,\text{crit}}$. As shown in Supp.\;Fig.\;\ref{fig:S7}b, the absolute value $|t_{\text{eff,L}}^{(2)}|$ vanishes for a relative phase of $\phi_{\text{r}}=0$, while $|t_{\text{eff,R}}^{(2)}|=0$ at $\phi_{\text{r}}=\pi$. This measurement confirms the second part of the relation in Eq.\;(\ref{tpmAgain}) and shows that the Dirac points appear at different points in the driving parameter space. In particular, even if double occupancies cannot be created by a density-assisted hopping process to the right when sitting at the point for which $|t_{\text{eff},\text{R}}^{(2)}|=0$, the amplitude for creating a double occupancy by tunnelling to the left is still finite. In general, we confirmed that the atoms tunnel with three different amplitudes and phases depending on the occupation of the involved lattice sites as expressed in the effective Hamiltonian in Eq.\;(\ref{HeffManySpin}) with the tunneling operators in Eqs.\;(\ref{TunnelOp}) and (\ref{GaugeOp}).

\end{document}